\documentclass[intlimits,twoside,a4paper]{article}

\usepackage{amsmath,amssymb}
\usepackage{graphicx}

\usepackage[T2A]{fontenc}
\usepackage[cp1251]{inputenc}

\usepackage{cmpj3}

\issue{2018}{21}{4}{43601}
\doinumber{10.5488/CMP.21.43601}

\title[Properties of t-X$_{3}$As$_{4}$ (X $=$ Si, Ge and Sn) by first-principles calculations]{The structural, mechanical, electronic, optical and thermodynamic properties of t-X$_{3}$As$_{4}$ (X $=$ Si, Ge and Sn) by first-principles calculations}
\author[R. Yang \textsl{et al.}]{R. Yang\refaddr{label1}, Y. Ma\refaddr{label1}, Q. Wei\refaddr{label1}, 
	D. Zhang\refaddr{label2}, Y. Zhou\refaddr{label3}}
\addresses{
\addr{label1} School of Physics and Optoelectronic Engineering, Xidian University, Xi'an, Shaanxi 710071, PR China
\addr{label2} National Supercomputing Center in Shenzhen, Shenzhen 518055, PR China
\addr{label3} Leihua Electronic and Technology Research Institute, Aviation Industry Corporation of China,\\ Wuxi, Jiangsu 214063, PR China
}

\date{Received July 3, 2018, in final form October 19, 2018}

\begin{document}

\maketitle

\begin{abstract}
The structural, mechanical, electronic, optical and thermodynamic properties 
of the t-X$_{\mathrm{3}}$As$_{\mathrm{4}}$ (X $=$ Si, Ge and Sn) with 
tetragonal structure have been investigated by first principles 
calculations. Our calculated results show that these compounds are 
mechanically and dynamically stable. By the study of elastic anisotropy, it 
is found that the anisotropic of the t-Sn$_{\mathrm{3}}$As$_{\mathrm{4}}$ is 
stronger than that of t-Si$_{\mathrm{3}}$As$_{\mathrm{4}}$ and 
t-Ge$_{\mathrm{3}}$As$_{\mathrm{4}}$. The band structures and density of 
states show that the t-X$_{\mathrm{3}}$As$_{\mathrm{4}}$ (Si, Ge and Sn) are 
semiconductors with narrow band gaps. Based on the analyses of electron 
density difference, in t-X$_{\mathrm{3}}$As$_{\mathrm{4}}$ As atoms get 
electrons, X atoms lose electrons. The calculated static dielectric 
constants, $\varepsilon_{1} (0)$, are 15.5, 20.0 and 15.1~eV for 
t-X$_{\mathrm{3}}$As$_{\mathrm{4}}$ (X $=$ Si, Ge and Sn), respectively. The 
Dulong-Petit limit of t-X$_{\mathrm{3}}$As$_{\mathrm{4}}$ is about 10~J~mol$^{\mathrm{-1}}$K$^{\mathrm{-1}}$. The thermodynamic stability successively decreases 
 from t-Si$_{\mathrm{3}}$As$_{\mathrm{4}}$ to 
t-Ge$_{\mathrm{3}}$As$_{\mathrm{4}}$ to t-Sn$_{\mathrm{3}}$As$_{\mathrm{4}}$.


\pacs 61.82.Bg, 62.20.dc, 71.20.Be, 71.15.Mb
\end{abstract}

\section{Introduction}

Liu and Cohen predict the $\beta $-C$_{\mathrm{3}}$N$_{\mathrm{4}}$ 
possesses the property of low compressibility. The structural and electronic 
properties of IV$_{\mathrm{3}}$V$_{\mathrm{4}}$~compounds~have attracted 
more and more attention \cite{1}. The hardness of the cubic 
silicon nitride was experimentally determined to be 35.31~GPa as the third 
hardest material after diamond and cubic boron nitride \cite{2}. 
The thermal stability of the cubic silicon nitride was studied by X-ray 
powder diffraction and scanning electron microscopy which shows that the 
material is stable up to 1673 K in air. So, the material is suitable for 
engineering superhard ceramics for high temperature structural applications 
\cite{2}. IV$_{\mathrm{3}}$V$_{\mathrm{4}}$ compounds have important potential 
applications in technical and scientific fields. The group IV nitrides have 
low compressibility and high hardness \cite{3}. They have a wide 
application prospect in cutting \cite{3}. Ching predicts the properties of the 
group IV nitrides using first principles theory \cite{1}. Feng investigates the 
properties of pseudocubic-X$_{\mathrm{3}}$P$_{\mathrm{4}}$ (X $=$ C, Si, Ge 
and Sn) by using the first principle calculations \cite{1}. They find that the modulus 
decreases with the atom change from C to Sn \cite{1}. 

The structural and electronic properties of 
pseudocubic-X$_{\mathrm{3}}$As$_{\mathrm{4}}$ are investigated by using 
first principles method \cite{1}. The 
pseudocubic-C$_{\mathrm{3}}$As$_{\mathrm{4}}$ is predicted to be metallic, 
pseudocubic-Si$_{\mathrm{3}}$As$_{\mathrm{4}}$, 
Ge$_{\mathrm{3}}$As$_{\mathrm{4}}$ and Sn$_{\mathrm{3}}$As$_{\mathrm{4}}$ 
are semiconductors \cite{4}. Although 
X$_{\mathrm{3}}$As$_{\mathrm{4}}$ have a wide range of applications in 
physics and chemistry, little research has been conducted regarding its mechanical and 
optical properties. Due to the lack of relevant experimental data at 
present, if we want to understand the application of tetragonal 
X$_{\mathrm{3}}$As$_{\mathrm{4}}$, further theoretical and computational 
research on its properties should be done.  

In this work, the structural parameters, mechanical, electronic, optical and 
thermodynamic properties of the tetragonal X$_{\mathrm{3}}$As$_{\mathrm{4}}$ 
are calculated by using the first-principles method based on plane waves and 
pseudo-potentials. By consulting the literature, we find that nobody has 
ever conducted experimental studies of the properties of the 
X$_{\mathrm{3}}$As$_{\mathrm{4}}$ (X $=$ Si, Ge and Sn). Therefore, we do not 
have the experimental data to compare with. 

\section{Computational methods}

The structural optimization, mechanical, electronic, optical and 
thermodynamic properties of t-X$_{3}$As$_{4}$ are calculated by using the 
density functional theory (DFT) with the generalized gradient approximation 
(GGA) parameterized by Perdew, Burke and Ernzerrof (PBE) in the CASTEP code 
\cite{5}. The band structures, electronic density of states 
(DOS) are calculated by PBE0 hybrid functional. The phonon spectra and 
phonon density of states (PHDOS) are calculated by using first-principles 
linear response method. Plane wave cut-off energy is set at 610~eV. The 
convergence test and energy cut-off analysis of $k$-point mesh samples show 
that the convergence of the Brillouin zone sampling and the kinetic energy 
cut-off are reliable and satisfy the computational requirements 
\cite{6,7}. The 
Broyden-Fletcher-Goldfarb-Shanno (BFGS) minimization scheme is used in 
geometry optimization. The value of self-consistent field tolerance 
threshold is set as $5.0\times 10^{-6}$~eV/atom. The maximum 
Hellmann-Feynman force, the maximum stress and the maximum displacement are 
set to be 0.01~eV/\AA, 0.02~GPa and $5.0\times 10^{-4}$~{\AA}  in geometry 
optimization, respectively. The Monkhorst-Pack $k$-points in the first 
irreducible Brillouin zone are given as $10\times 10\times 13$ for 
t-X$_{3}$As$_{4}$, respectively. The mechanical, electronic, optical and 
thermodynamic properties of t-X$_{3}$As$_{4}$ are computed according to 
the optimized crystal structures and parameters. 

\section{Results and discussions}
\subsection{Structural properties}

\begin{figure}[!b]
\centering
\includegraphics[height=3.7cm]{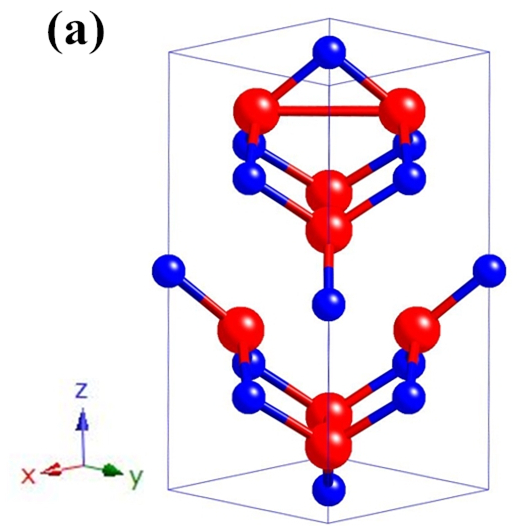} \quad
\includegraphics[height=3.7cm]{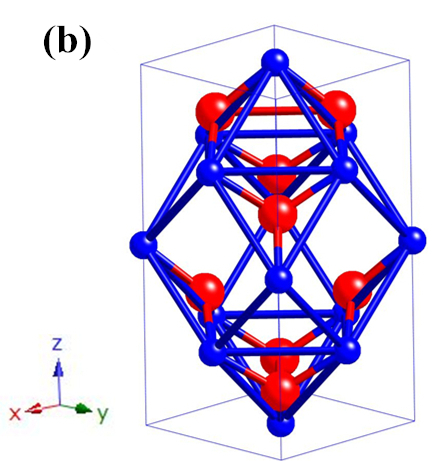} \quad
\includegraphics[height=1cm]{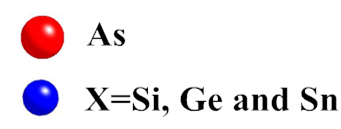}
\caption{(Colour online) Crystal structures of t-Si$_{3}$As$_{4}$, t-Ge$_{3}$As$_{4}$ and t-Sn$_{3}$As$_{4}$ (I-42M, No.~121).}
\label{fig1}
\end{figure}

\begin{table}[!b]
\centering
\caption{Lattice constants $a$, $b$, $c$ ({\AA}) and volume (\AA$^{3})$ of the t-X$_{3}$As$_{4}$.}
\vspace{2ex}
		\begin{tabular}{|c|c|c|c|c|c|c|c|}
			\hline\hline 
			Structure & $a$ & $b$ & $c$ & $\alpha $ & $\beta $ & $\gamma $ & $V$ \\ 
			\hline\hline 
			Si$_{3}$As$_{4}$&5.36&5.36&10.69&90.00&90.00&90.00&307.24\\
			\hline
			Ge$_{3}$As$_{4}$&5.48&5.48&11.00&90.00&90.00&90.00&330.56\\
			\hline
			Sn$_{3}$As$_{4}$&5.83&5.83&11.73&90.00&90.00&90.00&398.66\\
			\hline\hline
		      \end{tabular}
		\label{tab1}
\end{table}

The t-X$_{3}$As$_{4 }$ has a tetragonal structure with I-42M (No. 121). 
Figure~\ref{fig1} shows the crystal structures of t-X$_{3}$As$_{4}$ at 0~GPa. 
The t-Si$_{3}$As$_{4}$, t-Ge$_{3}$As$_{4}$ and t-Sn$_{3}$As$_{4}$ are all 
centro-symmetry structures (0, 0, 0) with 14~atoms/unit cell. The lattice 
parameters and volumes of t-X$_{3}$As$_{4}$ at 0~GPa are calculated on the 
basis of geometry optimization and are presented in table~\ref{tab1}. The 
density values of t-X$_{3}$As$_{4}$ are 4.15015~g/cm$^{3}$, 5.19889~g/cm$^{3}$ and 5.46287~g/cm$^{3}$, respectively.

\subsection{Mechanical properties}

The nine independent quantities of the elastic constants for 
t-X$_{3}$As$_{4}$ are calculated and presented in table~\ref{tab2}. The 
mechanical stabilities of t-X$_{3}$As$_{4}$ can be estimated by elastic 
constants. In this work, the mechanical stabilities and mechanical moduli of 
t-X$_{3}$As$_{4}$ are approximated by the corresponding relationships for a
tetragonal crystal class \cite{8,9,10,11}. The elastic stability criteria of tetragonal 
phases are given as follows \cite{12}:
\begin{equation}
\begin{array}{c}
\label{eq1}
C_{11} >0,\quad C_{33} >0, \quad C_{44} >0, \quad C_{66} >0, \\ 
(C_{11} -C_{12} )>0,\quad (C_{11} +C_{33} -2C_{13} )>0, \\ 
{2(C_{11} +C_{12} )+C_{33} +4C_{13} }>0. \\  
\end{array}
\end{equation}

As indicated in table~\ref{tab2}, the calculated elastic constants of the 
t-X$_{3}$As$_{4 }$ do satisfy the criteria, indicating that they are 
mechanically stable. The t-Si$_{3}$As$_{4}$ exhibits the largest elastic 
constants of $C_{11}$, $C_{22}$, $C_{33}$. For the t-Si$_{3}$As$_{4}$ and the 
t-Sn$_{3}$As$_{4}$, the calculated $C_{11}$ and $C_{22}$ are equal and they are 
larger than $C_{33}$. Hence, the mechanical strength in [100] and [010] 
directions is stronger than that in [001] direction. Moreover, $C_{44}$, $C_{55}$ 
and $C_{66}$ denote the shear moduli in (100), (010) and (001) crystal 
planes, respectively. For t-Ge$_{3}$As$_{4}$, the values of $C_{11}$ and 
$C_{22}$ are the same and smaller than $ C_{33}$. Hence, the mechanical strength 
in [100] and [010] directions are smaller than that in [001] direction. From 
table~\ref{tab2}, the $C_{44}$, $C_{55}$ of t-X$_{3}$As$_{4}$ are equal and they 
are larger than $C_{66}$. Therefore, the shear moduli at (100) and (010) 
crystal planes are larger than that of (001) crystal planes. 

\begin{table}[!b]
\centering
\caption{ Calculated independent elastic constants of the 
	t-X$_{3}$As$_{4}$ (X $=$ Si, Ge and Sn).}
	\vspace{2ex}
		\begin{tabular}{|c|c|c|c|c|c|c|c|c|c|}
			\hline\hline 
		Species&$C_{11}$& $C_{12}$& $C_{13}$& $C_{22}$& $C_{23}$& $C_{33}$& $C_{44}$& $C_{55}$& $C_{66}$\\
			\hline\hline
			Si$_{3}$As$_{4}$&97.34&35.86&36.85&97.34&36.85&94.60&49.34&49.34&42.90\\
			\hline
			Ge$_{3}$As$_{4}$&80.97&30.70&34.87&80.97&34.87&85.04&39.06&39.06&32.60\\
			\hline
			Sn$_{3}$As$_{4}$&54.68&24.29&27.06&54.68&27.06&52.30&25.16&25.16&21.49\\
			\hline\hline
		\end{tabular}
		\label{tab2}
\end{table}

From the calculated elastic constants, other mechanical parameters such as 
bulk modulus ($B)$, shear modulus ($G)$, Young's modulus ($E)$ and Poisson's ratio 
($v)$ can be derived using Voigt-Reuss-Hill (VRH) approximation \cite{8}. For 
t-X$_{3}$As$_{4}$, based on elastic constants, the Reuss shear modulus 
($G_\text{R})$ and the Voigt shear modulus ($G_\text{V})$ are as follows:
\begin{equation}
\begin{array}{c}
G_\text{R} =15/[4(S_{11} +S_{22} +S_{33} )-4(S_{12} +S_{13} +S_{23} 
	)+3(S_{44} +S_{55} +S_{66} )],
\end{array}
\end{equation}
\begin{equation}
\begin{array}{c}
G_\text{V} =(C_{11} +C_{22} +C_{33} -C_{12} -C_{13} -C_{23} 
)/15+(C_{44} +C_{55} +C_{66})/5.
\end{array}
\end{equation}

The Reuss bulk modulus ($B_\text{R}$) and the Voigt bulk modulus ($B_\text{V}$) are defined 
as
\begin{equation}
\begin{array}{c}
B_\text{R} ={1}/[(S_{11} +S_{22} +S_{33} )+2(S_{12} +S_{13} +S_{23} )],
\end{array}
\label{4}
\end{equation}
\begin{equation}
\begin{array}{c}
B_\text{V} =(C_{11} +C_{22} +C_{33} )/9+2(C_{12} +C_{13} +C_{23} )/9,
\end{array}
\end{equation}
where the relationship between $S_{ij}$ and $C_{ij}$ is shown as
\begin{equation}
\begin{array}{c}
[S_{ij} ]=[C_{ij} ]^{-1}.
\end{array}
\end{equation}

In the above formulae, the $C_{ij}$ represents the elastic stiffness matrix 
and the $S_{ij}$ represents the elastic flexibility matrix. The Hill's averages are taken from the averages of the two \cite{8}
\begin{equation}
\begin{array}{c}
B=(B_\text{V} +B_\text{R} )/2,
\end{array}
\end{equation}
\begin{equation}
\begin{array}{c}
G=(G_\text{V} +G_\text{R} )/2.
\end{array}
\end{equation}

The Young's modulus, $E$, and Poisson's ratio, $v$, can be 
	calculated by the equations
\begin{equation}
\begin{array}{c}
E=9BG/(3B+G),
\end{array}
\end{equation}
\begin{equation}
\begin{array}{c}
v=(3B-2G)/[2(3B+G)].
\end{array}
\label{10}
\end{equation}

From (\ref{4}) to (\ref{10}), the calculated physical quantities are 
	presented in the table~\ref{tab3}.

\begin{table}[!t]
\caption{Calculated mechanical moduli of t-X$_{3}$As$_{4}$, 
	including bulk modulus ($B_\text{V}$, $B_\text{R}$ and $B)$, shear modulus ($G_\text{V}$, 
	$G_\text{R}$ and $G)$, Young's modulus ($E$) and Poisson's ratio ($v$), $B/G $ and $H_\text{V}$, in GPa.}
	\vspace{2ex}
	\begin{center}
		\begin{tabular}{|c|c|c|c|c|c|c|c|c|c|}
			\hline \hline
			Species& $B_\text{V}$& $B_\text{R}$& $G_\text{V}$& $G_\text{R}$& $B$& $G$& $E$& $v$& $B/G$\\
			\hline \hline
			Si$_{3}$As$_{4}$&56.49&56.48&40.30&38.27&56.49&39.28&95.67&0.22&1.44\\
			\hline
			Ge$_{3}$As$_{4}$&49.76&49.66&31.91&30.50&49.71&31.20&77.42&0.24&1.59\\
			\hline
			Sn$_{3}$As$_{4}$&35.38&35.38&19.91&18.42&35.38&19.16&48.70&0.27&1.85\\
			\hline\hline
		\end{tabular}
	\end{center}
	\label{tab3}
\end{table}

From the calculated results, t-Si$_{3}$As$_{4}$ has the largest value of 
bulk modulus among t-X$_{3}$As$_{4}$, which indicates that 
t-Si$_{3}$As$_{4}$ has a lower compressibility. The shear modulus of 
t-Si$_{3}$As$_{4}$ is the largest in t-X$_{3}$As$_{4}$, indicating that 
t-Si$_{3}$As$_{4}$ has a strong rigidity. The Young's modulus values of 
t-X$_{3}$As$_{4}$ show that t-Si$_{3}$As$_{4}$ has a larger hardness.

The Poisson's ratio represents the stability of the shear strain of the 
crystal. For a typical metal, the value should be 0.33. For the 
ionic-covalent crystal, the value ranges from 0.2 to 0.3. Poisson's ratio of 
the strong covalent crystal is relatively small, usually below 0.15 \cite{8}. For 
t-X$_{3}$As$_{4}$, the Poisson's ratios show that they are ionic-covalent 
crystal.

The ratio $B/G$ reflects the brittleness or ductility of a material, and the 
critical value is close to 1.75 \cite{13}. Below 1.75, the material shows 
brittleness; otherwise it shows toughness. As indicated in table~\ref{tab3}, 
the $B$/$G$ values for t-Si$_{3}$As$_{4}$ and t-Ge$_{3}$As$_{4}$ are smaller than 
1.75, and the $B$/$G$ value for t-Sn$_{3}$As$_{4}$ is higher than 1.75. Hence, 
t-Si$_{3}$As$_{4}$ and t-Ge$_{3}$As$_{4}$ show brittleness and 
t-Sn$_{3}$As$_{4}$ shows toughness.

It is well known that elastic anisotropy plays an important role in 
engineering science and crystal physics. The shear anisotropic factor for 
the (100) shear planes between [011] and [010] directions can be written by \cite{14}: 
\begin{equation}
\begin{array}{c}
A_1 ={4C_{44} }/(C_{11}+C_{33}-2C_{13}).
\end{array}
\end{equation}
For the (010) shear plane between [101] and [001] direction is:
\begin{equation}
\begin{array}{c}
A_2 ={4C_{55} }/(C_{22} +C_{33} -2C_{23} ).
\end{array}
\end{equation}
For the (001) shear planes between [110] and [010] direction is:
\begin{equation}
\begin{array}{c}
A_3 ={4C_{66} }/(C_{11} +C_{22} -2C_{12}).
\end{array}
\end{equation}

For isotropic crystal, the factors $ A_{1}, A_{2}$ and $A_{3}$ must be 1, while 
the deviation from 1 is a measure of the degree of the elastic anisotropy. 
Furthermore, since the compound is tetragonal, rather than cubic, the shear 
anisotropic factors are not sufficient to describe the elastic anisotropy. 
Therefore, the anisotropy of the linear bulk modulus should be considered. 
The anisotropy of the bulk modulus along $a$ and $c$ axes relative to the anisotropy 
along $b$ axis can be estimated using the following equations:
\begin{equation}
A_{Ba} ={B_a }/{B_b }\,, \quad A_{Bc} ={B_c }/{B_b }.
\end{equation}

When the value is 1, it represents an elastic isotropy, but if it is not equal to 1, it is an
elastic anisotropy. Where $B_{a}$, $B_{b}$ and $B_{c}$ are the bulk moduli 
along different crystal axes, defined as
\begin{equation}
B_a =a({\rd P}/{\rd a})={\Lambda }/(1+\alpha +\beta ),
\end{equation}
\begin{equation}
B_b =b({\rd P}/{\rd b})={B_a }/{\alpha },
\end{equation}
\begin{equation}
B_c =c({\rd P}/{\rd c})={B_a }/{\beta },
\end{equation}
\begin{equation}
\Lambda =C_{11} +2C_{12} \alpha +C_{22} \alpha ^2+2C_{13} \beta +C_{33} 
\beta ^2+2C_{23} \alpha \beta ,
\end{equation}
\begin{equation}
\alpha =\frac{(C_{11} -C_{12} )(C_{33} -C_{13} )-(C_{23} -C_{13} )(C_{11} 
	-C_{13} )}{(C_{33} -C_{13} )(C_{22} -C_{12} )-(C_{13} -C_{23} )(C{ 
	}_{12}-C_{23} )}\,,
\end{equation}
\begin{equation}
\beta =\frac{(C_{22} -C_{12} )(C_{11} -C_{13} )-(C_{11} -C_{12} )(C_{23} 
	-C_{12} )}{(C_{22} -C_{12} )(C_{33} -C_{13} )-(C_{12} -C_{23} )(C_{13} 
	-C_{23} )}.
\end{equation}

In addition, the percentage of elastic anisotropy for bulk modulus $A_{B}$ and 
shear modulus $A_{G}$ in polycrystalline materials can be used as follows:
\begin{align}
	 A_B &=(B_\text{V} -B_\text{R} )/(B_\text{V} +B_\text{R} ),\\ 
	 A_G &=(G_\text{V} -G_\text{R} )/(G_\text{V} +G_\text{R} ). 
\end{align}

The implication of the definition is that the value of zero corresponds to 
elastic isotropy and the value of 100{\%} identifies the largest elastic 
anisotropy.

\begin{table}[!t]
\caption{ The shear anisotropic factors $A_{1}$, $A_{2}$, $A_{3}$, 
compressibility anisotropy factors $A_{Ba}$, $A_{Bc}$, percentage anisotropy 
in compressibility $A_{B}$ and shear $A_{G}$, bulk modulus along the 
tetragonal crystallographic axes $a$, $b$, $c$ ($B_{a}$, $B_{b}$ and $B_{c}$, in Gpa) 
in tetragonal I-42M(121) space group.}
\vspace{2ex}
	\begin{center}
		\begin{tabular}{|c|c|c|c|c|c|c|c|c|c|c|}
			\hline \hline
			 & $A_{1}$& $A_{2}$& $A_{3}$& $A_{Ba}$& $A_{Bc}$& $A_{B}$& $A_{G}$& $B_{a}$& $B_{b}$& $B_{c}$\\
			\hline \hline
			Si$_{3}$As$_{4}$&1.67&1.67&1.40&1.00&1.00&0.00&0.03&171.17&171.17&166.12\\
			\hline
			Ge$_{3}$As$_{4}$&1.62&1.62&1.30&1.00&1.20&0.00&0.03&140.82&140.82&168.47\\
			\hline
			Sn$_{3}$As$_{4}$&1.90&1.90&1.41&1.00&1.02&0.00&0.04&105.62&105.62&107.23\\
			\hline\hline
		\end{tabular}
	\end{center}
	\label{tab4}
	\vspace{-3mm}
\end{table}

The calculated results are listed in table~\ref{tab4}. It is seen that 
t-X$_{3}$As$_{4}$ are elastic anisotropic. The t-Sn$_{3}$As$_{4}$ shows to be more 
anisotropic than t-Si$_{3}$As$_{4}$ and t-Ge$_{3}$As$_{4}$. For 
t-X$_{3}$As$_{4}$, the bulk moduli along $a$ axis are equal to that along the $b$. For t-Si$_{3}$As$_{4}$, the bulk modulus along $c$ axis is smaller than 
that along $a$ axis and $b$ axis. For t-Ge$_{3}$As$_{4}$ and t-Sn$_{3}$As$_{4}$, 
the bulk moduli along $c$ axis are larger than that along $a$ axis and $b$ axis. 
In addition, we also noticed that the percentage bulk moduli anisotropy 
$A_{B}$ is smaller than shear modulus anisotropy $A_{G}$, suggesting that the 
structures are anisotropic in compressibility.

\subsection{Phonon spectra}

\begin{figure}[t]
\centering
		\includegraphics[width=5cm]{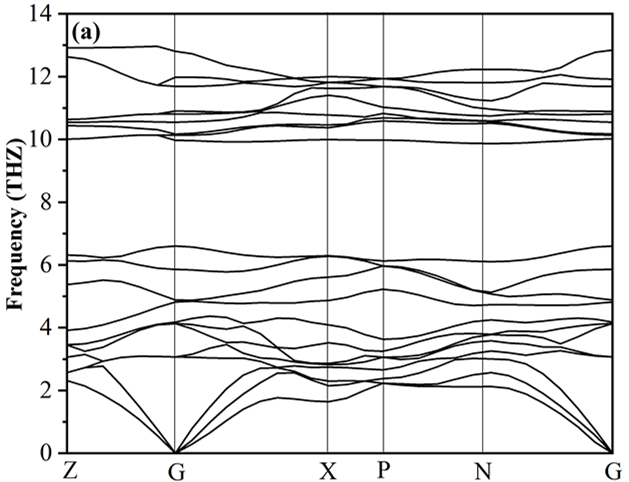}\qquad
		\includegraphics[width=5cm]{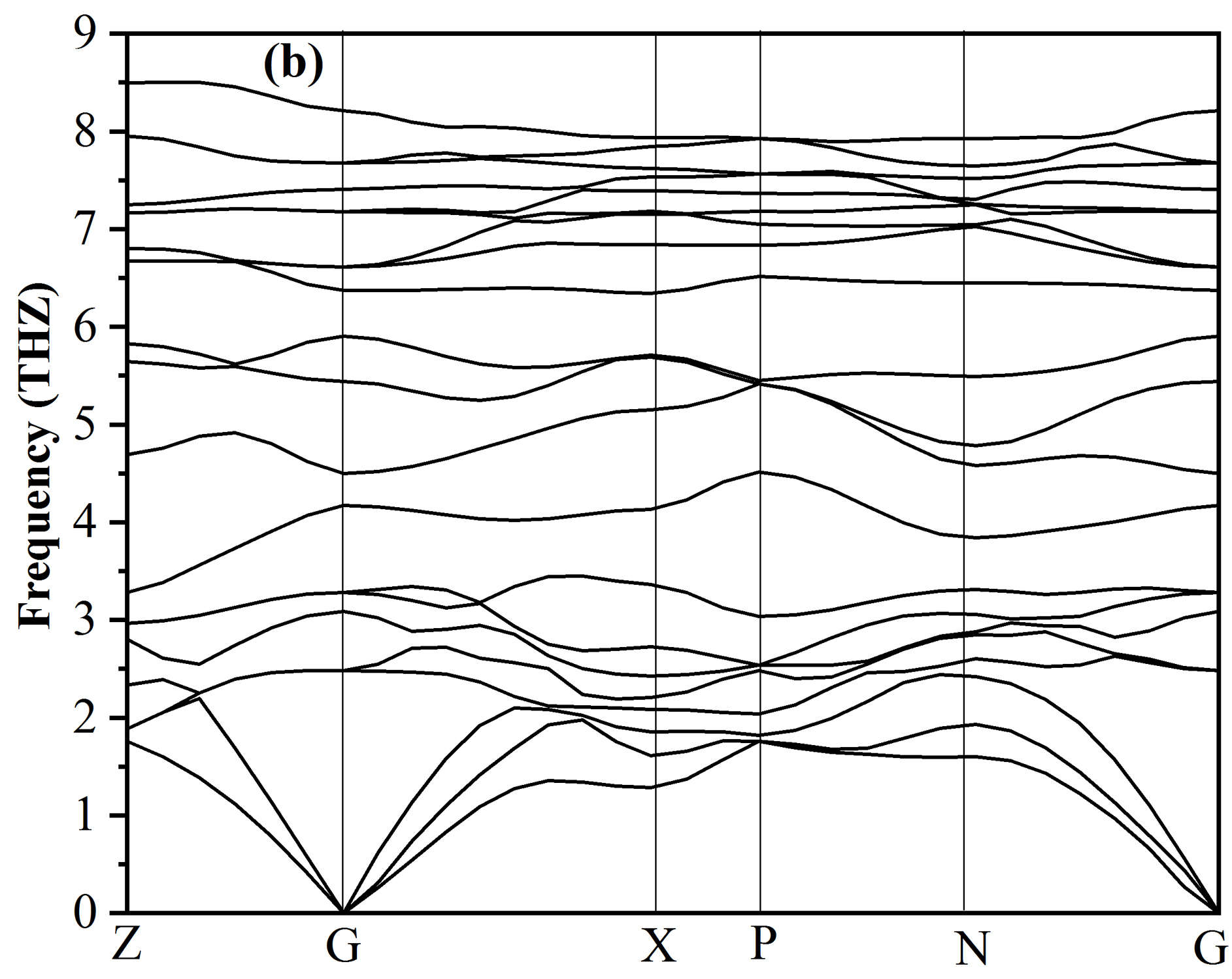}
		\includegraphics[width=5cm]{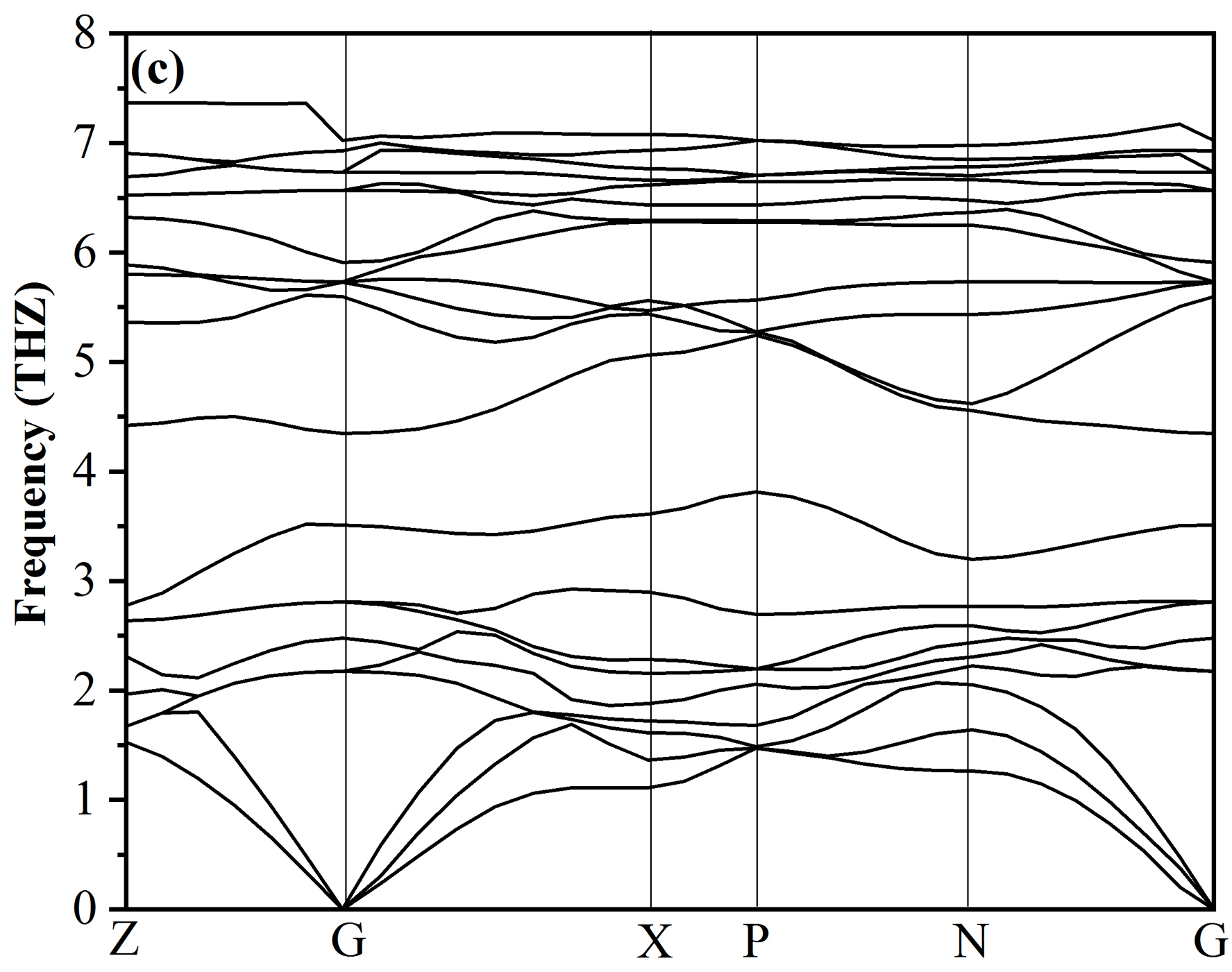}
\caption{ The phonon band structures (a) t-Si$_{3}$As$_{4}$, (b) t-Ge$_{3}$As$_{4}$, (c) t-Sn$_{3}$As$_{4}$.}
\label{fig2}
\end{figure}

\begin{figure}[!t]
\centering
		\includegraphics[width=5cm]{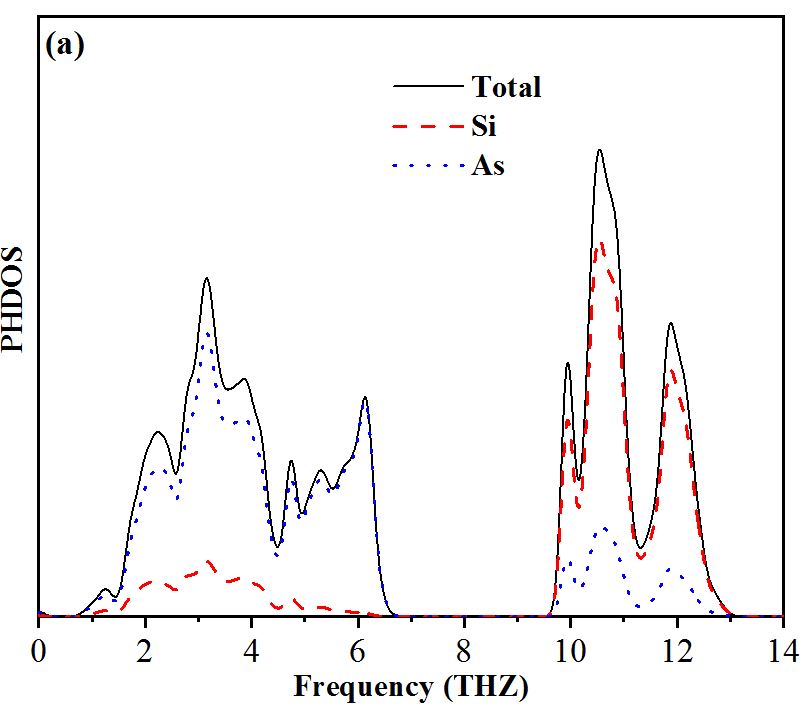} \qquad
		\includegraphics[width=5cm]{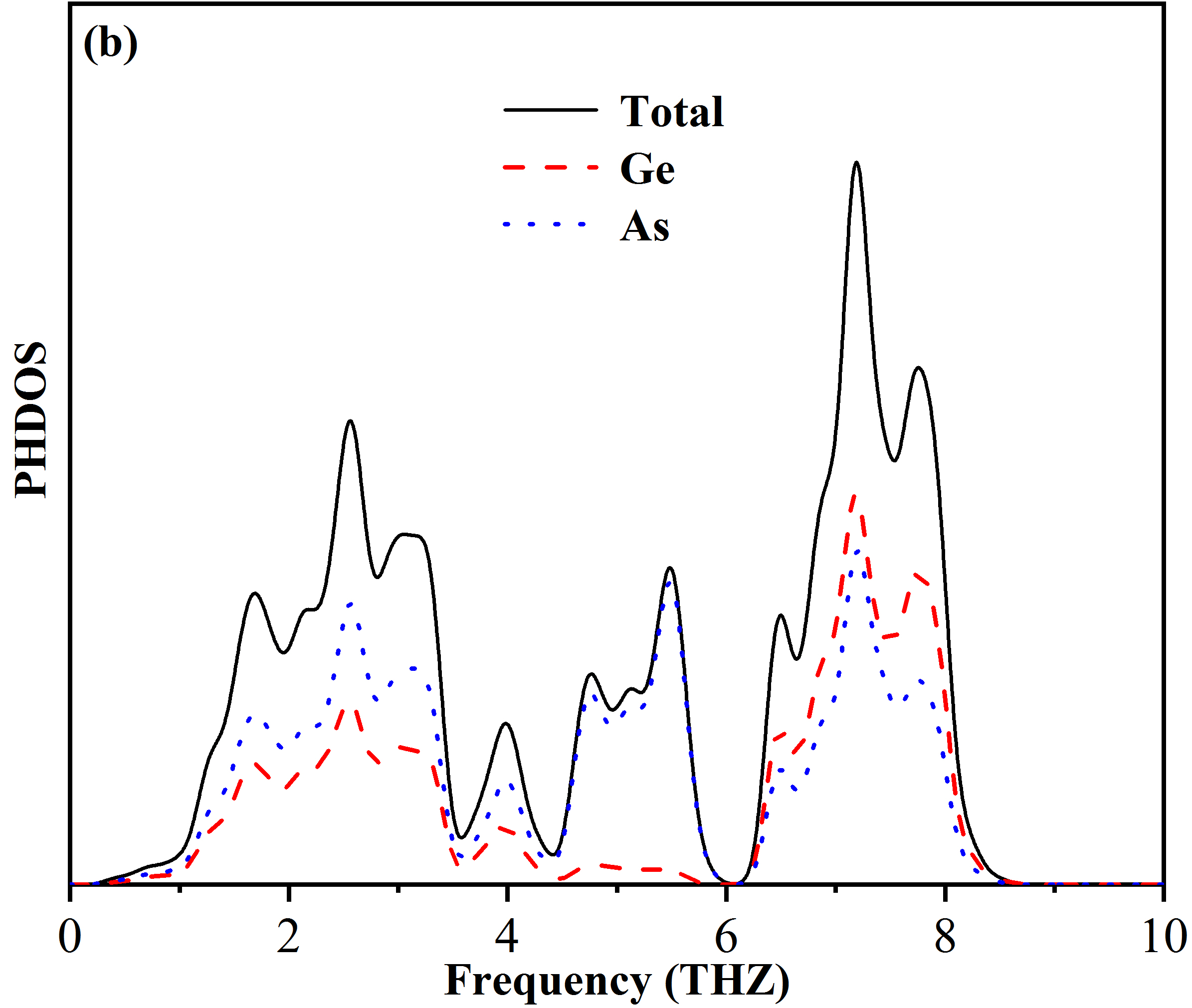}
		\includegraphics[width=5cm]{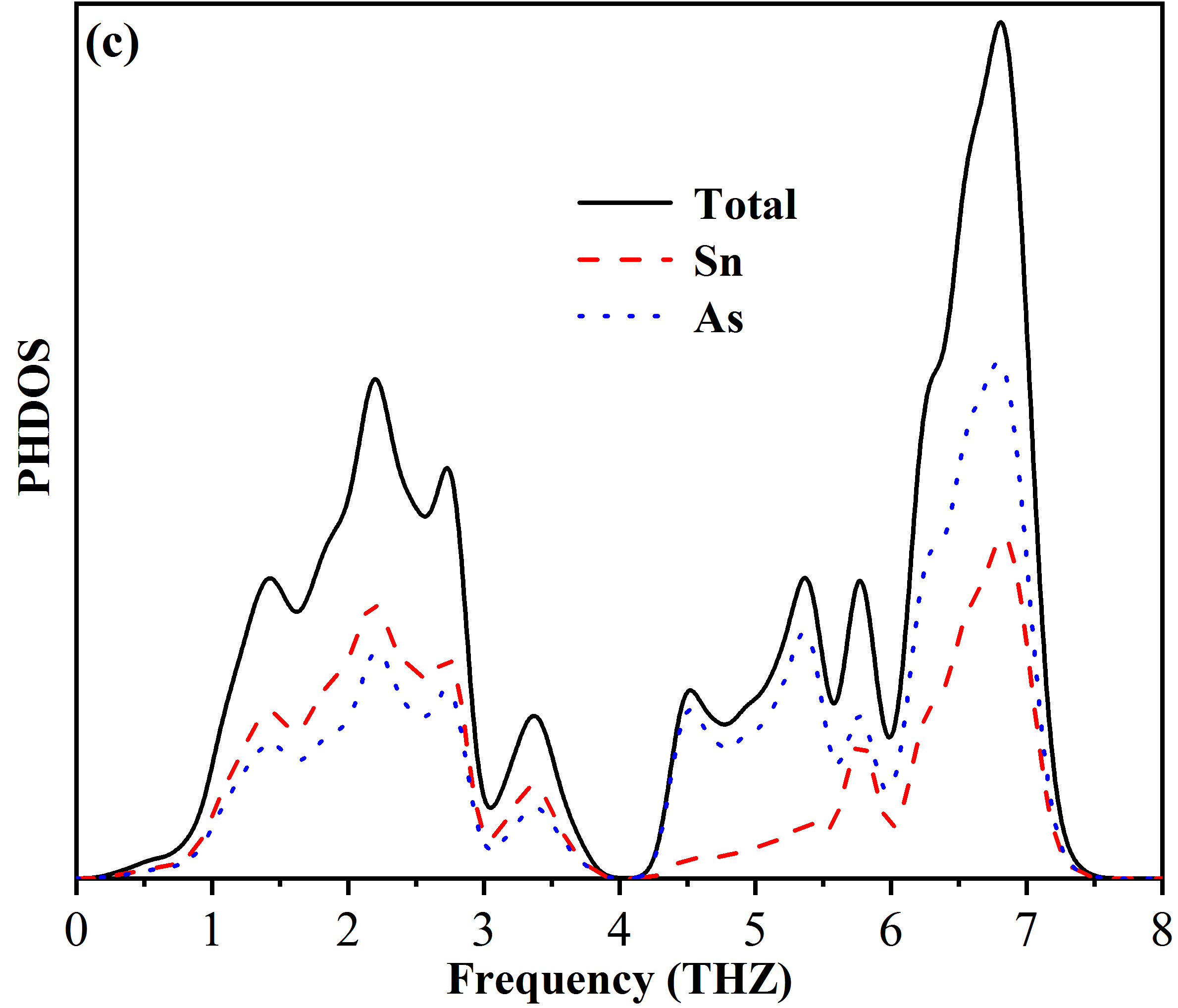}
\caption{(Colour online) The phonon density of states (PHDOS) (a) t-Si$_{3}$As$_{4}$, (b) 
		t-Ge$_{3}$As$_{4}$ and (c) t-Sn$_{3}$As$_{4}$.}
\label{fig3}
\end{figure}

The calculated phonon dispersion spectra of t-X$_{3}$As$_{4}$ at 0~GPa are 
shown in figure~\ref{fig2}. No any imaginary phonon frequencies in the entire 
Brillouin zone confirms that t-X$_{3}$As$_{4}$ are dynamically stable \cite{15,16}. 
Since there are seven atoms in the protocell of t-X$_{3}$As$_{4}$, the 
crystal should have twenty one dispersion relationships in theory. That is, 
there should be twenty one phonon spectra calculation curves, which shows 
that the calculated results in this paper are consistent with the 
theoretical conclusions. Some of the twenty one phonon spectral curves 
overlap in some range of wave vectors. There are three acoustic branches and 
eighteen optical branches. Lattice vibrational modes play a major role in 
Raman scattering and infrared absorption at point $\Gamma $. From 
figure~\ref{fig2}~(a) and figure~\ref{fig3}~(a), we can see that an optical gap 
exists in the phonon dispersion spectrum of t-Si$_{3}$As$_{4}$. This is due 
to the large quality difference between Si atom and As atom. From 
figure~\ref{fig2}~(b) and figure~\ref{fig3}~(b), we can observe that the optical 
band gap of t-Ge$_{3}$As$_{4}$ is smaller than that of t-Si$_{3}$As$_{4}$, 
which is due to the mass difference between Ge and As being smaller. A 
similar situation occurs in t-Sn$_{3}$As$_{4}$.

\looseness=-1 The phonon density of states (PHDOS) of t-X$_{3}$As$_{4}$ is calculated 
by CASTEP and shown in figure~\ref{fig3}, respectively. For 
t-Si$_{3}$As$_{4}$, we can observe that in the range of $0.8{-}6.8$~eV, the 
total PHDOS is mainly from As, which indicates that the vibration of As 
atoms is dominant in this frequency band. In the range of $9.5{-}13.0$~eV, the 
total PHDOS is mainly from Si, which indicates that the vibration of Si 
atoms is dominant in this frequency band. For t-Ge$_{3}$As$_{4}$, in the 
range of $0.2{-}6.1$~eV, the total PHDOS is mainly from As, which indicates that 
the vibration of As is dominant in this frequency band. In the range of 
$6.2{-}8.8$~eV, the total PHDOS is from Ge and As, which shows that Ge and As 
have a similar vibrational probability. For t-Sn$_{3}$As$_{4}$, in the 
range of $0.2{-}4.0$~eV, the total PHDOS is from Sn and As, which shows that Sn 
and As have a similar vibrational probability. And in the range of $4.2{-}7.6$~eV, the total PHDOS is mainly from As and partial from Sn, which indicates 
that the vibration of As is dominant in this frequency band.

\subsection{Band structures and densities of states}

The electronic properties of t-X$_{3}$As$_{4}$ are analyzed at 0~GPa. The 
DOS is calculated to have a further insight into the bonding characteristics 
of t-X$_{3}$As$_{4}$. 

\begin{figure}[!t]
\centering
		\includegraphics[width=6cm]{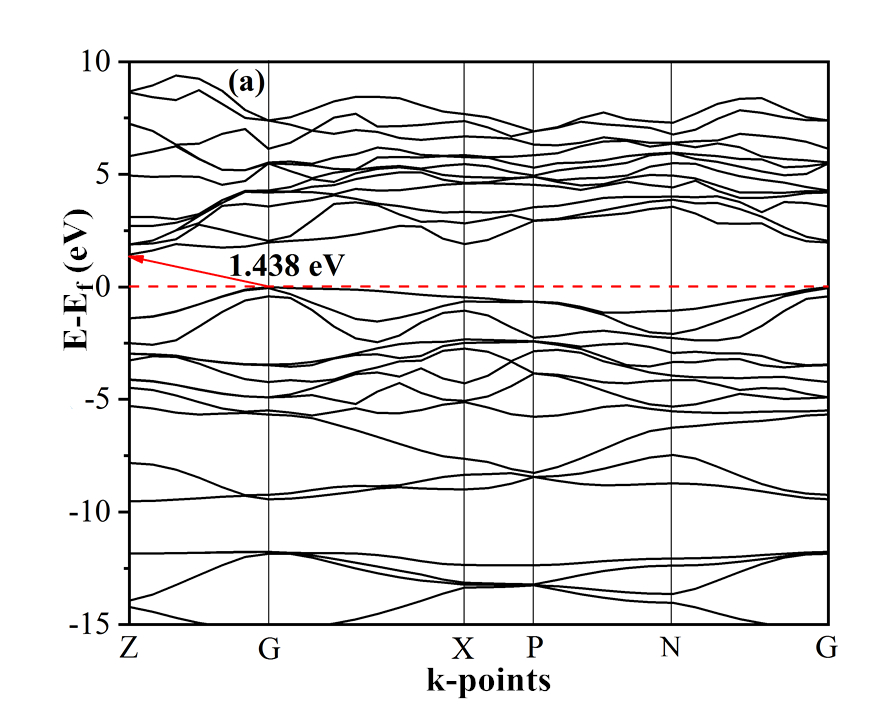} \qquad
		\includegraphics[width=6cm]{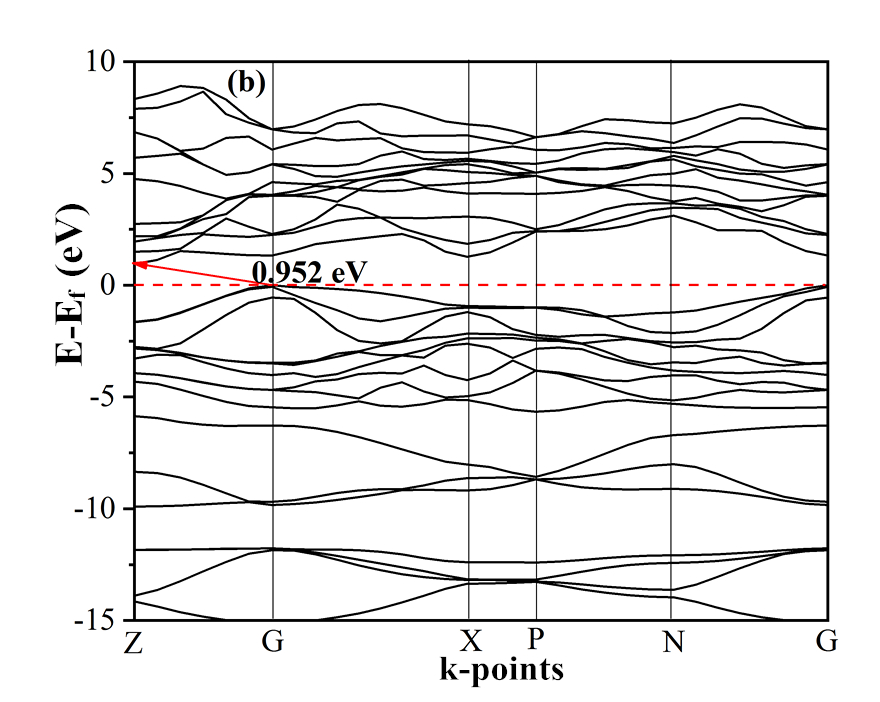}
		\includegraphics[width=6cm]{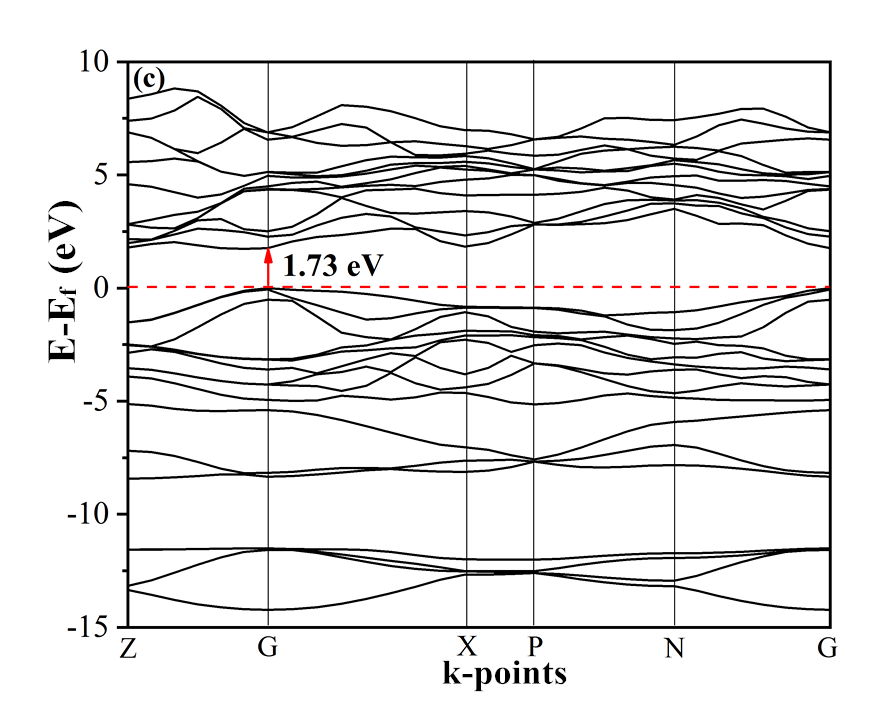}
\caption{(Colour online) The band structures (a) t-Si$_{3}$As$_{4}$, (b) 
		t-Ge$_{3}$As$_{4}$, (c) t-Sn$_{3}$As$_{4}$.}
	\label{fig4}
\end{figure}
\begin{figure}[!t]
\centering
		\includegraphics[width=6cm]{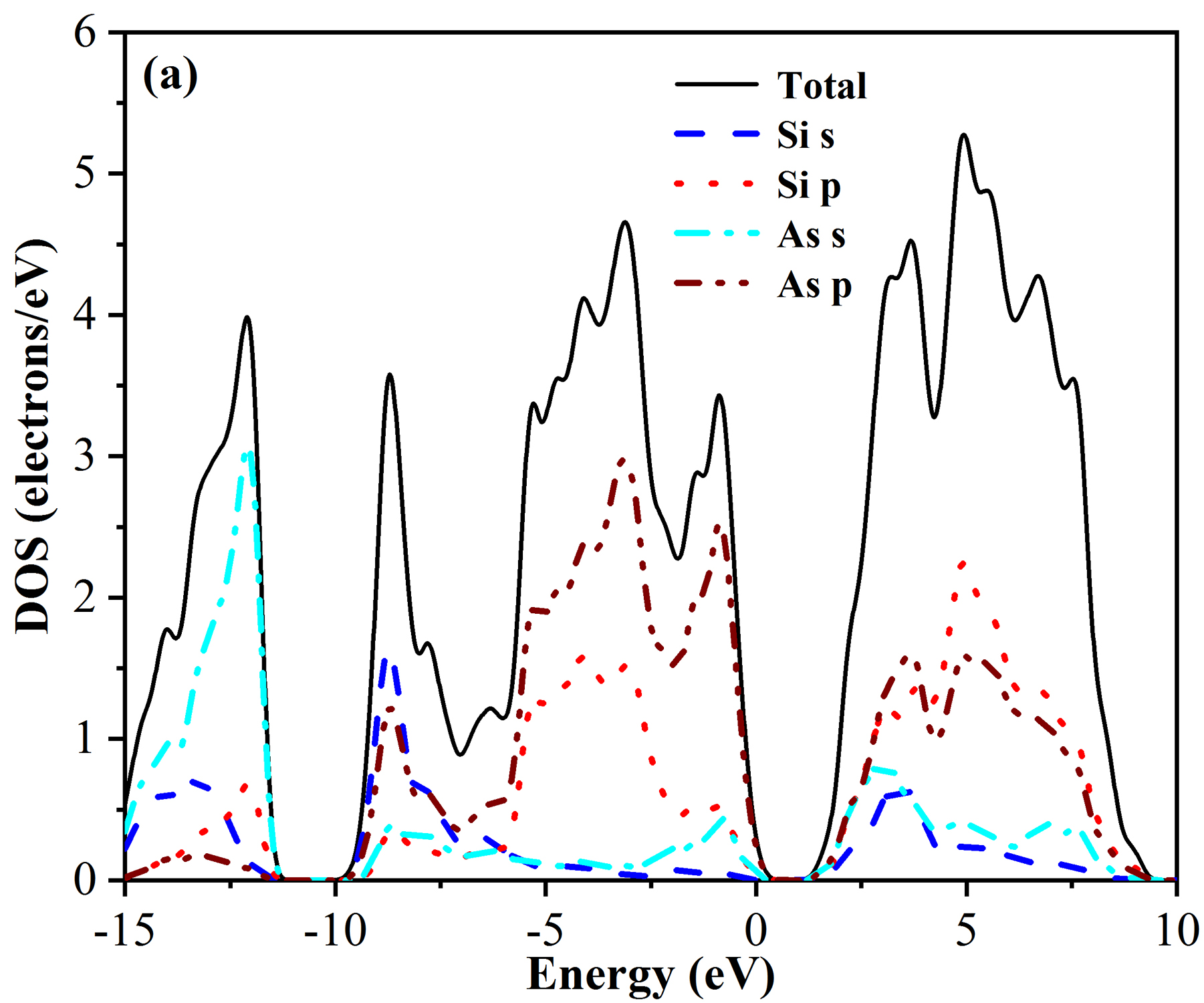} \qquad
		\includegraphics[width=6cm]{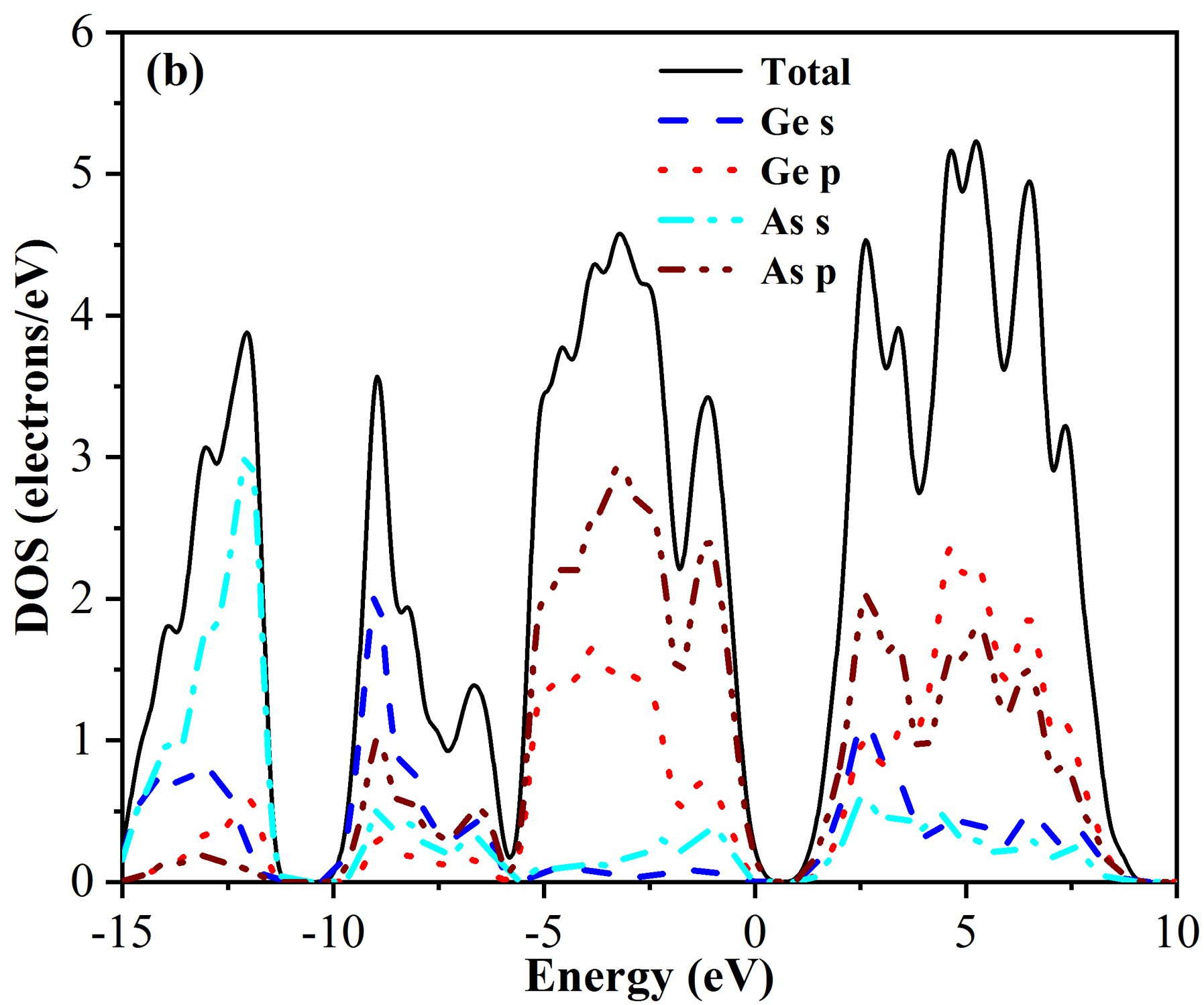}
		\includegraphics[width=6cm]{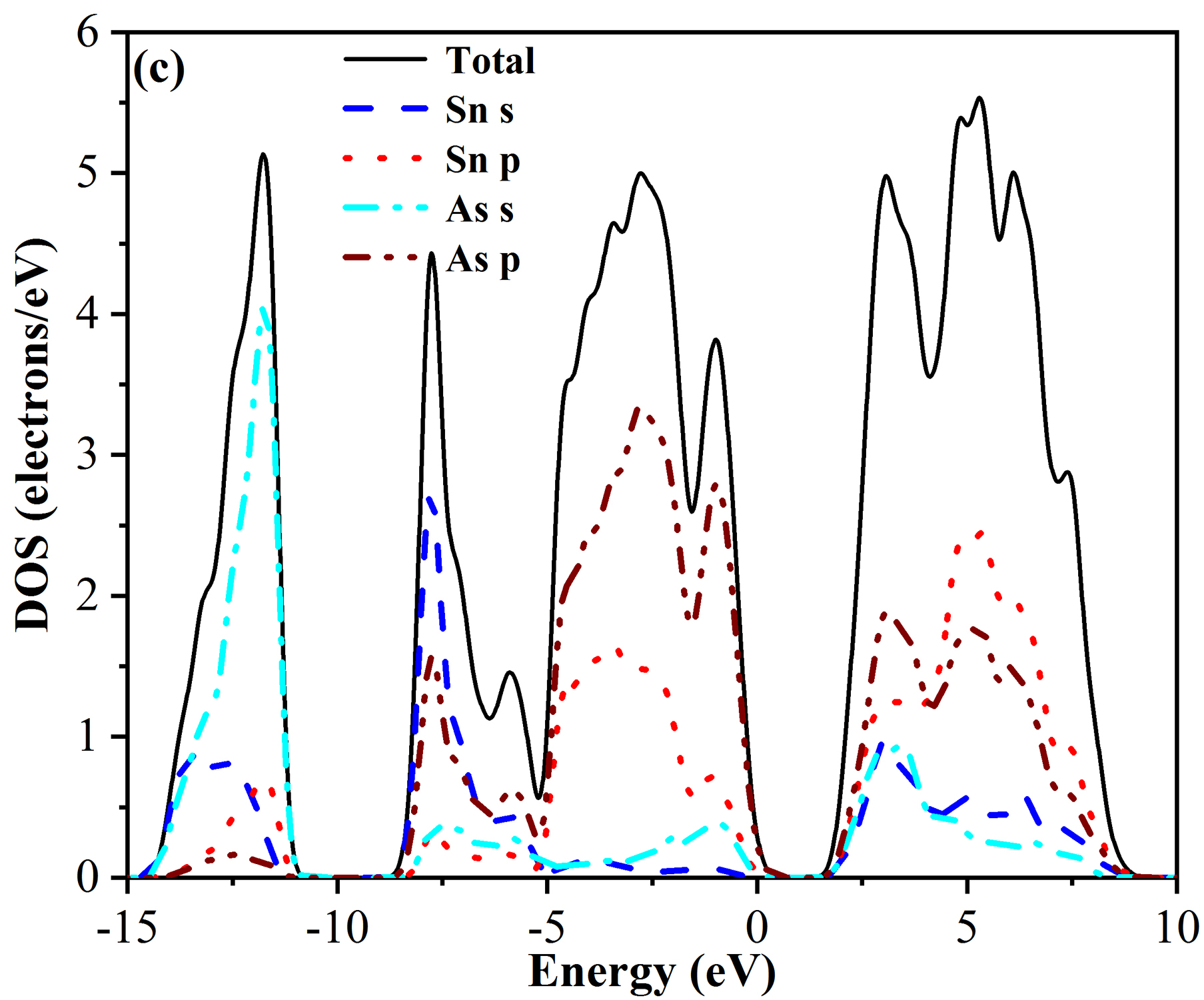}
\caption{(Colour online) The electronic densities of states (a) t-Si$_{3}$As$_{4}$, 
		(b) t-Ge$_{3}$As$_{4}$, (c) t-Sn$_{3}$As$_{4}$.}
	\label{fig5}
\end{figure}

As indicated in figure~\ref{fig4}, for t-Si$_{3}$As$_{4}$, the valence maximum 
is at high symmetric $\Gamma $-point and the conduction band minimum is at Z-point, 
which indicates that t-Si$_{3}$As$_{4}$ is an indirect band semiconductor with a 
band gap of 1.438~eV. For t-Ge$_{3}$As$_{4}$, the valence maximum is at high 
symmetric $\Gamma $-point and the conduction band minimum is at Z-point, which 
indicates that t-Ge$_{3}$As$_{4}$ is an indirect band semiconductor with a band 
gap of 0.952~eV. For t-Sn$_{3}$As$_{4}$, the highest point of the valence 
band and the lowest point of the conduction band are at the $\Gamma $ point. It 
shows that t-Sn$_{3}$As$_{4}$ is a direct band semiconductor with a band gap 
of 1.73~eV.

The density of states of t-X$_{3}$As$_{4}$ are presented in figure~\ref{fig5}. For t-Si$_{3}$As$_{4}$, the main bonding peaks are in the range of 
$-15{-}10$~eV. The total DOS in the valence band mainly comes from As $s$ and As 
$p$, with partially from Si $s$ and Si $p$. Si $p$ and As $p$ contribute most to the 
total DOS in the conduction band, with partially from As $s$ and Si $s$. For 
t-Ge$_{3}$As$_{4}$, the total DOS in the valence band mainly comes from As 
$s$, As $p$ and Ge $s$, with partially from Ge $p$. In the conduction band, the 
total DOS mainly come from Ge~$p$ and As $p$ with partially from Ge $s$ and As $s$. 
For t-Sn$_{3}$As$_{4}$, the total DOS in the valence band mainly comes from 
As $s$, As $p$ and Sn $s$, with partially from Sn $p$. In the conduction band, the 
total DOS mainly come from Sn $p$ and As $p$ with partially from Sn $s$ and As $s$. 

\subsection{Electron density difference and Mulliken charge population}

\begin{figure}[!b]
\centering
		\includegraphics[width=3.8cm]{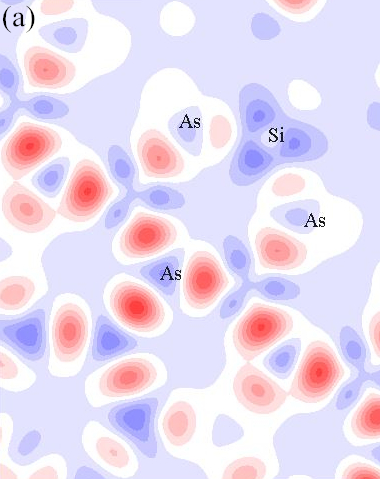}~~
		\includegraphics[width=3.8cm]{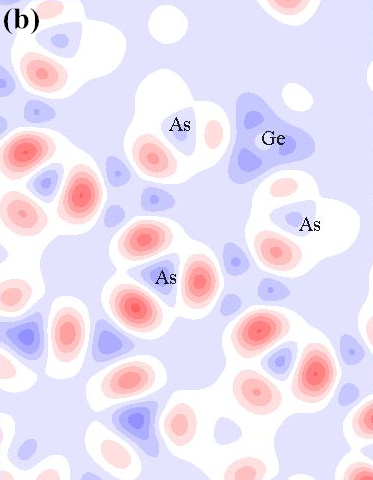}~~
		\includegraphics[width=3.8cm]{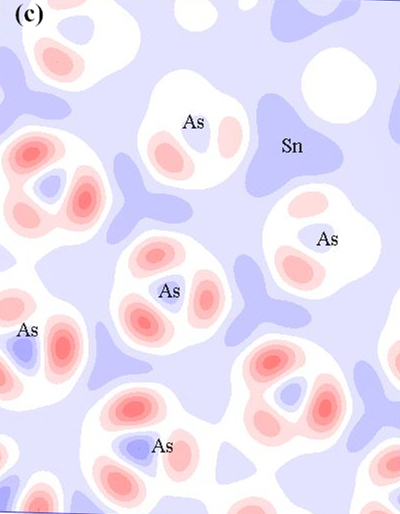}~~
		\includegraphics[width=2.8cm]{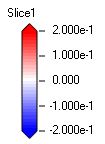}
\caption{(Colour online) Electron density difference distribution for (a) 
		t-Si$_{3}$As$_{4}$, (b) t-Ge$_{3}$As$_{4}$, (c) t-Sn$_{3}$As$_{4}$.}
\label{fig6}
\end{figure}

In order to investigate the chemical bonding, the electron density 
differences are calculated and the results are shown in figure~\ref{fig6}. The 
electron density differences are the discrepancy between the electron 
densities of the total system and the unperturbed electron densities of X 
(Si, Ge and Sn) and As \cite{16}. The contour plots show the electron density 
differences (between $-0.2$ and $0.2$) due to the formation of chemical bonds in 
the I-42M lattice, which are relative to the electron density in an isolated 
atom. It is helpful to analyze how chemical bonds are formed. The electron 
density differences are useful for identifying the types of chemical bonds. 
From figure~\ref{fig6}, we can see that As atoms get electrons, X (Si, Ge and 
Sn) lose electrons. The electrovalent bond is formed between X (Si, Ge and 
Sn) atoms and As atoms. The combination of X and As atoms is dependent on 
the ionic effect of Coulomb attraction. In order to further investigate the bonding 
behaviour of the three materials, we also obtain the Mulliken charge 
population. The Mulliken population results are given in table~\ref{tab5}. 
Table~\ref{tab5} gives a quantitative determination of the number of loss 
electrons per X (Si, Ge and Sn) atom, the number of acquired electrons per X 
(Si, Ge and Sn) and the charger of per X atom and per As atom after the 
formation of Si$_{3}$As$_{4}$. From table~\ref{tab5}, it is shown that the 
bonding behaviour of t-X$_{3}$As$_{4}$ is a combination of covalent and ionic 
nature.

\begin{table}[!t]
\caption{ The calculated atomic Mulliken charges (e) for 
	t-X$_{3}$As$_{4}$ (X $=$ Si, Ge and Sn).}
	\begin{center}
		\begin{tabular}{|c|c|c|c|c|c|}
			\hline \hline 
			Structure& Species& $s$& $p$& Total& Charge (e)\\
			\hline \hline 
			Si$_{3}$As$_{4}$&Si(1)&1.42&2.42&3.83&0.17\\
			 &Si(2)&1.43&2.39&3.82&0.18\\
			 &Si(3)&1.43&2.39&3.82&0.18\\
			 &As(1)&1.63&3.50&5.13&$-$0.13\\
			 &As(2)&1.63&3.50&5.13&$-$0.13\\
			 &As(3)&1.63&3.50&5.13&$-$0.13\\
			 &As(4)&1.63&3.50&5.13&$-$0.13\\
			\hline
			Ge$_{3}$As$_{4}$&Ge(1)&1.56&2.39&3.95&0.05\\
			 &Ge(2)&1.56&2.37&3.93&0.07\\
			 &Ge(3)&1.56&2.37&3.93&0.07\\
			 &As(1)&1.67&3.38&5.04&$-$0.04\\
			 &As(2)&1.67&3.38&5.04&$-$0.04\\
			 &As(3)&1.67&3.38&5.04&$-$0.04\\
			 &As(4)&1.67&3.38&5.04&$-$0.04\\
			\hline
			Sn$_{3}$As$_{4}$&Sn(1)&1.55&2.28&3.82&0.18\\
			&Sn(2)&1.53&2.26&3.79&0.21\\
			&Sn(3)&1.53&2.26&3.79&0.21\\
			&As(1)&1.60&3.55&5.15&$-$0.15\\
			&As(2)&1.60&3.55&5.15&$-$0.15\\
			&As(3)&1.60&3.55&5.15&$-$0.15\\
			&As(4)&1.60&3.55&5.15&$-$0.15\\
			\hline\hline 
		\end{tabular}
	\end{center}
	\label{tab5}
\end{table}

\subsection{Optical properties}

\begin{figure}[!t]
	\centering
		\includegraphics[width=2.9in]{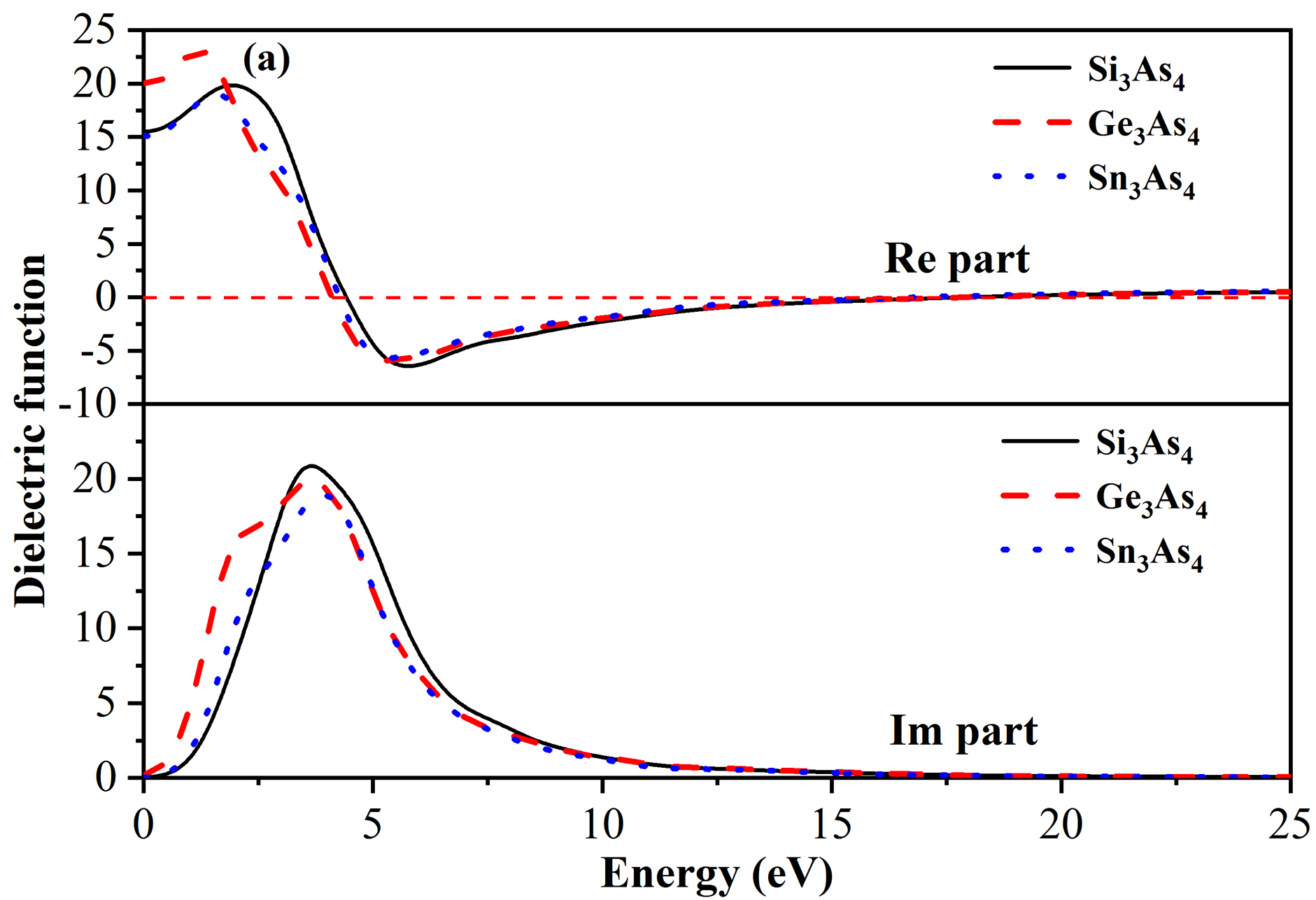}
		\includegraphics[width=2.9in]{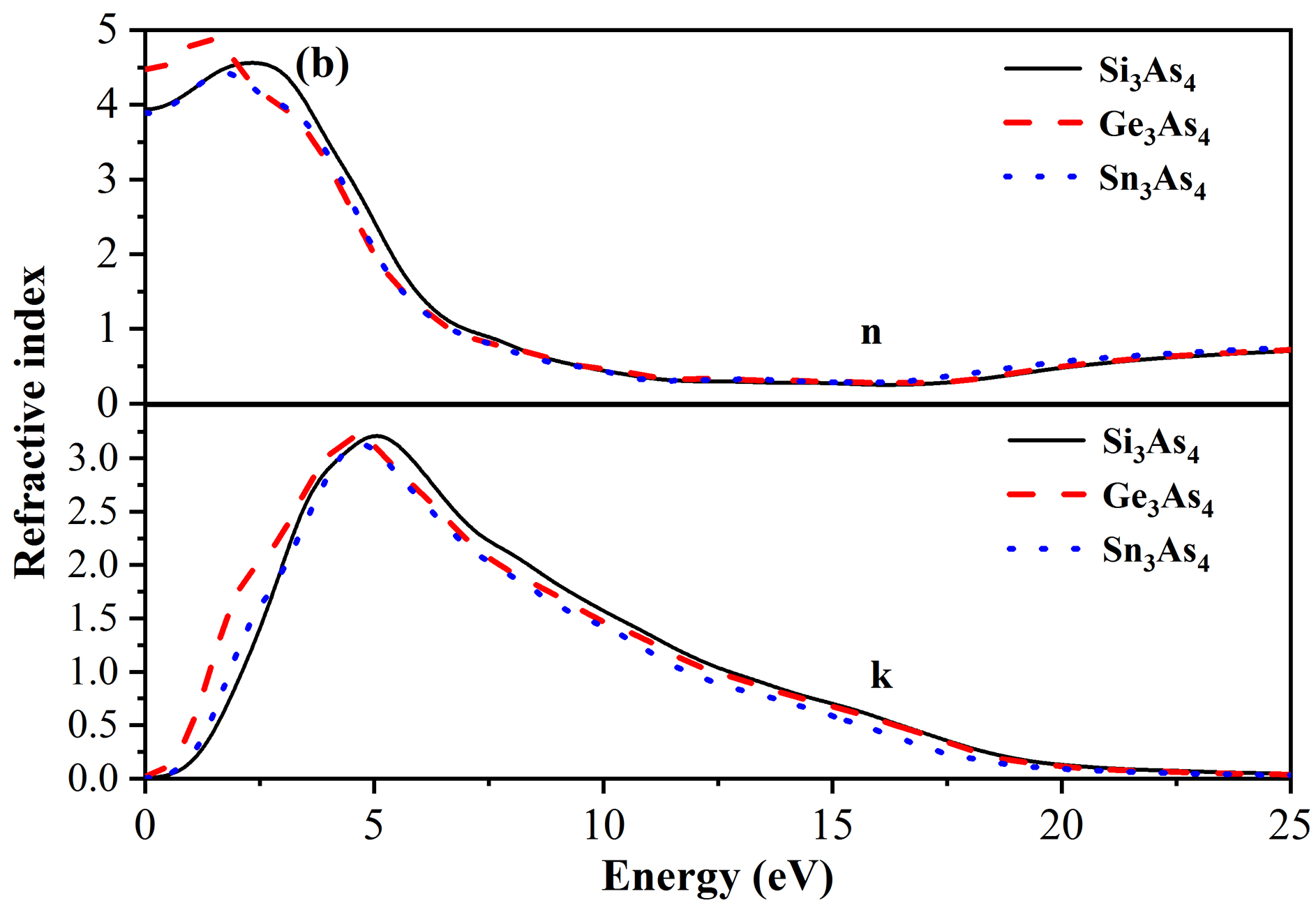}
		~\\
		\hspace{-2ex}
		\includegraphics[width=2.9in]{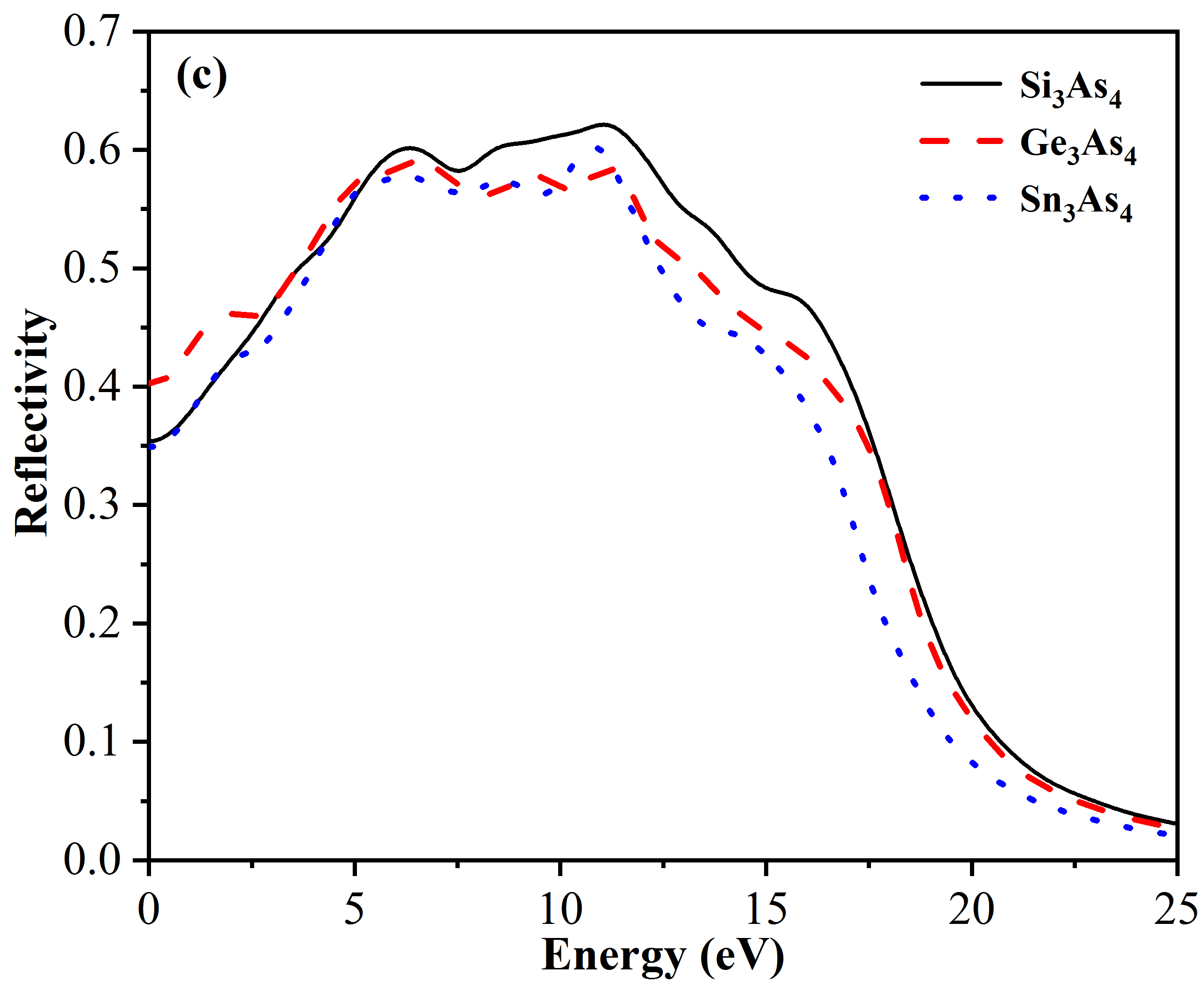}
		\includegraphics[width=2.9in]{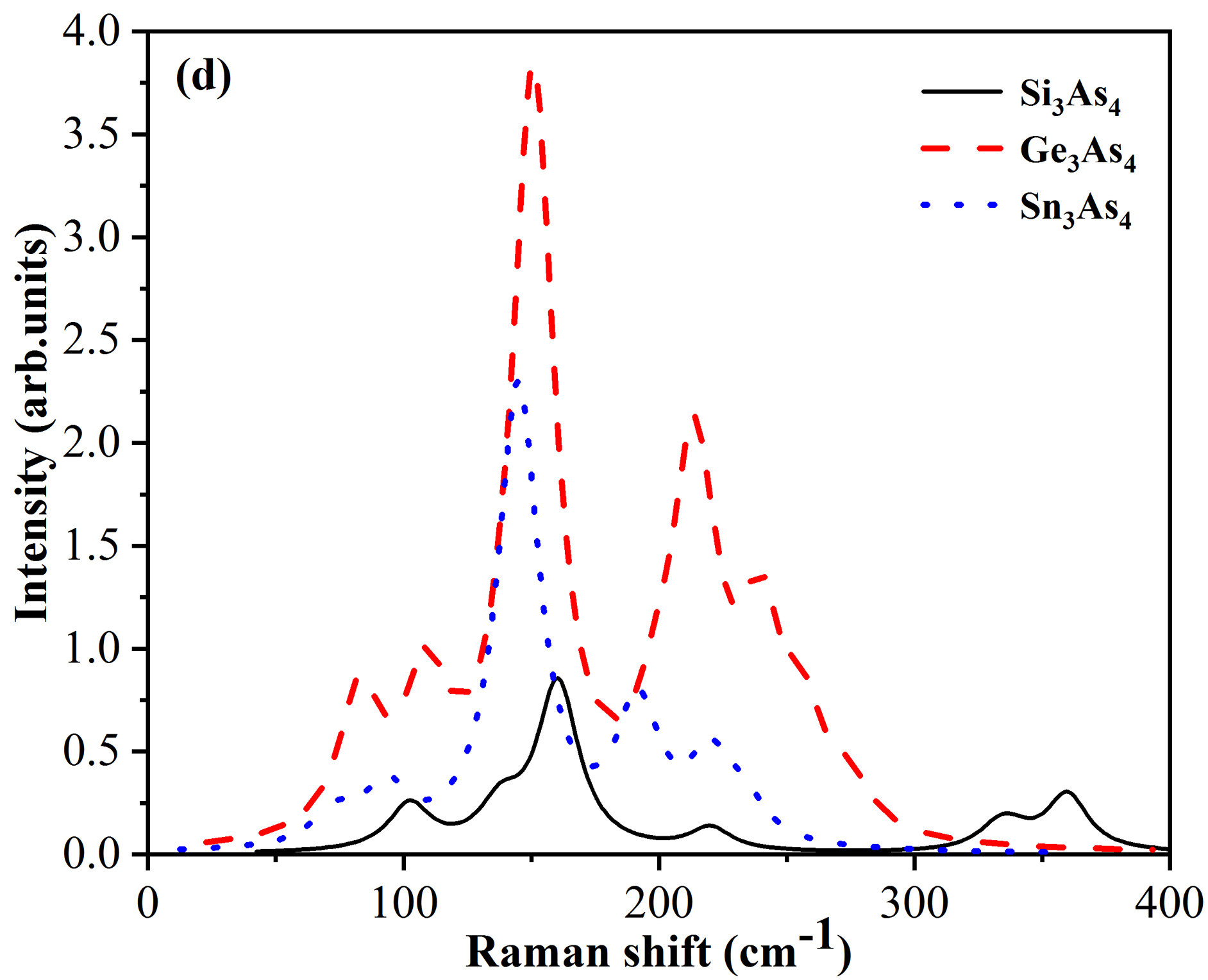}
\caption{(Colour online) (a) Real and imaginary parts of dielectric function, (b) 
		real and imaginary parts of refractive index, (c) optical reflectivity 
		spectrum, (d) Raman intensities for the t-X$_{3}$As$_{4}$ (X $=$ Si, Ge and Sn).}
\label{fig7}
\end{figure}

Dielectric function is the most common characteristic of materials. It can 
characterize the response of materials to incident electromagnetic waves 
\cite{17}. Optical properties of materials can usually be evaluated on the basis 
of a complex dielectric function, which depends on the frequency. Complex 
dielectric function is shown as follows \cite{18}:
\begin{equation}
\varepsilon (\omega )=\varepsilon _1 (\omega )+\ri\varepsilon _2 (\omega ),
\end{equation}
where $\varepsilon _1 (\omega )$ is the real part, $\varepsilon _2 (\omega 
)$ is the imaginary part. The real and imaginary parts of the dielectric 
function are given in figure~\ref{fig7}~(a). The calculated static dielectric 
constants, $\varepsilon _1 (0)$, are 15.5, 20.0 and 15.1~eV for 
t-Si$_{3}$As$_{4}$, t-Ge$_{3}$As$_{4}$ and t-Sn$_{3}$As$_{4}$, respectively. For t-Si$_{3}$As$_{4}$, t-Ge$_{3}$As$_{4}$ and t-Sn$_{3}$As$_{4}$, the real parts of 
dielectric function enhance with an increasing photon energy and get to the 
highest values at about 1.94, 1.30 and 1.48~eV, respectively. The imaginary 
part curves increase with an increase of photon energy and get to the 
highest values at about 3.69, 3.54 and 3.97~eV.

Refractive index $n(\omega )$ and extinction coefficient $k(\omega )$ of t-X$_{3}$As$_{4}$ can be obtained from $\varepsilon _1 (\omega )$ and $\varepsilon _2 (\omega )$: 
\begin{equation}
\begin{split}
n(\omega )=\left\{\frac{1}{2}\Big[\varepsilon_{1} (\omega )+\sqrt {\varepsilon_{1} 
	^{2}(\omega )+\varepsilon_{2}^{2}(\omega )} \Big]\right\}^{1/2},
\end{split}
\label{23}
\end{equation}
\begin{equation}
\begin{split}
k(\varepsilon )=\left\{\frac{1}{2}\Big[-\varepsilon_{1} (\omega )+\sqrt {\varepsilon_{1} 
	^{2}(\omega )+\varepsilon_{2}^{2}(\omega )} \Big]\right\}^{1/2}.
\end{split}
\label{24}
\end{equation}

The static refractive indices are found to be 
3.94, 4.48 and 3.89~eV for t-Si$_{3}$As$_{4}$, t-Ge$_{3}$As$_{4}$ and t-Sn$_{3}$As$_{4}$, respectively. The values of 
$n(\omega)$ increase with an increasing photon energy in the visible light 
region, and in the ultraviolet band, reach the peaks at about 2.24, 1.40 and 
1.67~eV for t-Si$_{3}$As$_{4}$, t-Ge$_{3}$As$_{4}$ and t-Sn$_{3}$As$_{4}$, respectively. In figure~\ref{fig7}~(c), the 
reflective coefficients are displayed. The main peaks lie at 11.1, 6.55 and 
10.8~eV, respectively.

Vibrational Raman spectroscopy is one of the widely used optical techniques 
in materials science. It is a standard method for quality control of a 
production line. It is very effective in determining the occurrence of new 
phases or structural changes at extreme conditions. Moreover, it can be used 
in the absence of a long-range structural order as for liquid or amorphous 
materials. Raman spectroscopy can link Raman lines to specific 
microstructures. In order to further verify the crystal structure, the 
micro-Raman spectra of t-Si$_{3}$As$_{4}$, t-Ge$_{3}$As$_{4}$ and t-Sn$_{3}$As$_{4}$ were calculated by first-principles. Our Raman intensity 
results excited by the 514.5~nm incident light are presented in figure~\ref{fig7}~(d). The Raman line shape is assumed to be Lorentzian, and the line-width 
is fixed at 10~cm$^{-1}$ FWHM. The peaks mainly locate at 103, 161, 220, 
336, and 361~cm$^{-1}$ for t-Si$_{3}$As$_{4}$, 83.8, 109, 151, 214 and 239~cm$^{-1}$ for t-Ge$_{3}$As$_{4}$, 92.8, 145, 191 and 219~cm$^{-1}$ for 
t-Sn$_{3}$As$_{4}$.

\subsection{Thermodynamic properties}

For t-X$_{3}$As$_{4}$, we have studied the thermodynamic properties based on 
the phonon density of states at 0--1000~K temperature. According to CASTEP, 
the heat capacity is contributed by the lattice, $C_{V}$ is
\begin{equation}
\begin{split}
C_V (t)=k\int {\frac{\big(\frac{\hbar \omega }{kT}\big)^2\exp \big(\frac{\hbar \omega 
		}{kT}\big)}{\big[\exp \big(\frac{\hbar \omega }{kT}\big)-1\big]^2}} F(\omega )\rd\omega. 
\end{split}
\end{equation}

\begin{figure}[!t]
\centering
\includegraphics[width=2.8in]{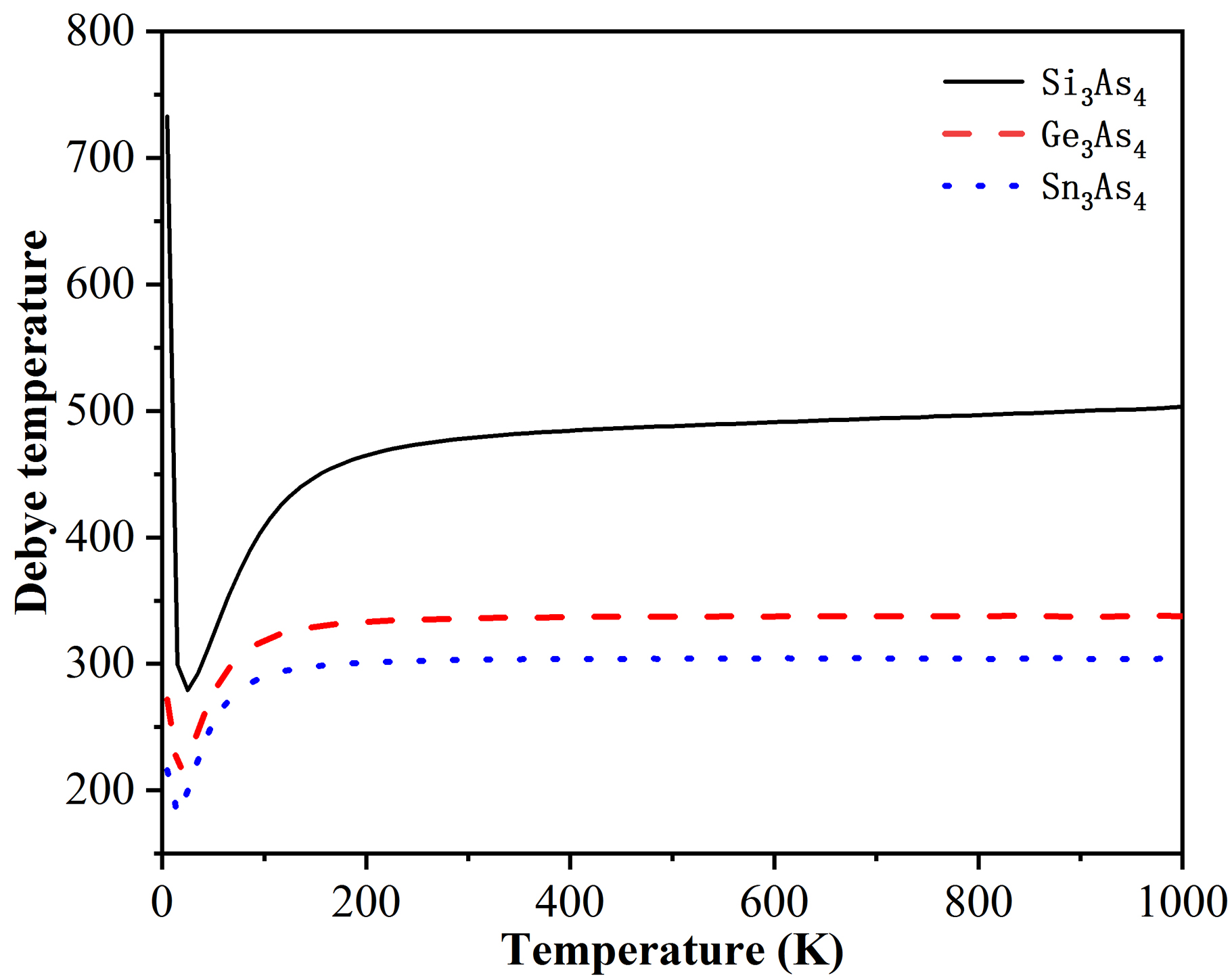}	
\caption{(Colour online) The relationship between Debye temperature and temperature 
for t-X$_{3}$As$_{4}$.}
\label{fig8}
\end{figure}

Debye temperature can be used to measure the properties of crystals, such as 
melting temperature, elastic constants and specific heat \cite{19}. With the 
development of cryogenic technology, the deviation between Debye theory and 
practice increases. It is shown that the Debye temperature is different 
at different temperatures. According to the temperature dependence of Debye 
temperature at a constant volume, some general theoretical predictions can be 
made \cite{20}. Heat capacity in Debye model is given by
\begin{equation}
\begin{split}
C_V^\text{D} (T)=9Nk\left(\frac{T}{\theta _\text{D} }\right)^3\int_0^{\theta _\text{D} /T} 
{\frac{x^4\re^x}{(\re^x-1)^2}} \rd x,
\end{split}
\end{equation}
where $N$ is the number of atoms per cell. Thus, the value of the Debye 
temperature, $\theta _\text{D}$, at a certain temperature, $T$, is obtained by calculating 
the actual heat capacity according to equation~(\ref{23}), then by inverting 
equation~(\ref{24}). The relations of the Debye temperature with temperature 
are given as figure~\ref{fig8}. The Debye temperature of t-Si$_{3}$As$_{4}$ is 
larger than that of t-Ge$_{3}$As$_{4}$ and t-Sn$_{3}$As$_{4}$ between 0 and 
1000~K. It is found that the Debye temperature decreases with an increase of 
temperature, and then increases after reaching a minimum, and that minimum 
values are 279, 216 and 181~K for t-Si$_{3}$As$_{4}$, t-Ge$_{3}$As$_{4}$ and t-Sn$_{3}$As$_{4}$. While above 300~K, 
$\theta _\text{D}$ of t-Si$_{3}$As$_{4}$, t-Ge$_{3}$As$_{4}$ and t-Sn$_{3}$As$_{4}$ show weak temperature dependence and approach the 
constant values. At 1000~K, values of $\theta _\text{D}$ are 503~K, 338~K and 304~K, 
respectively.

\begin{figure}[!t]
\centering
\includegraphics[width=2.8in]{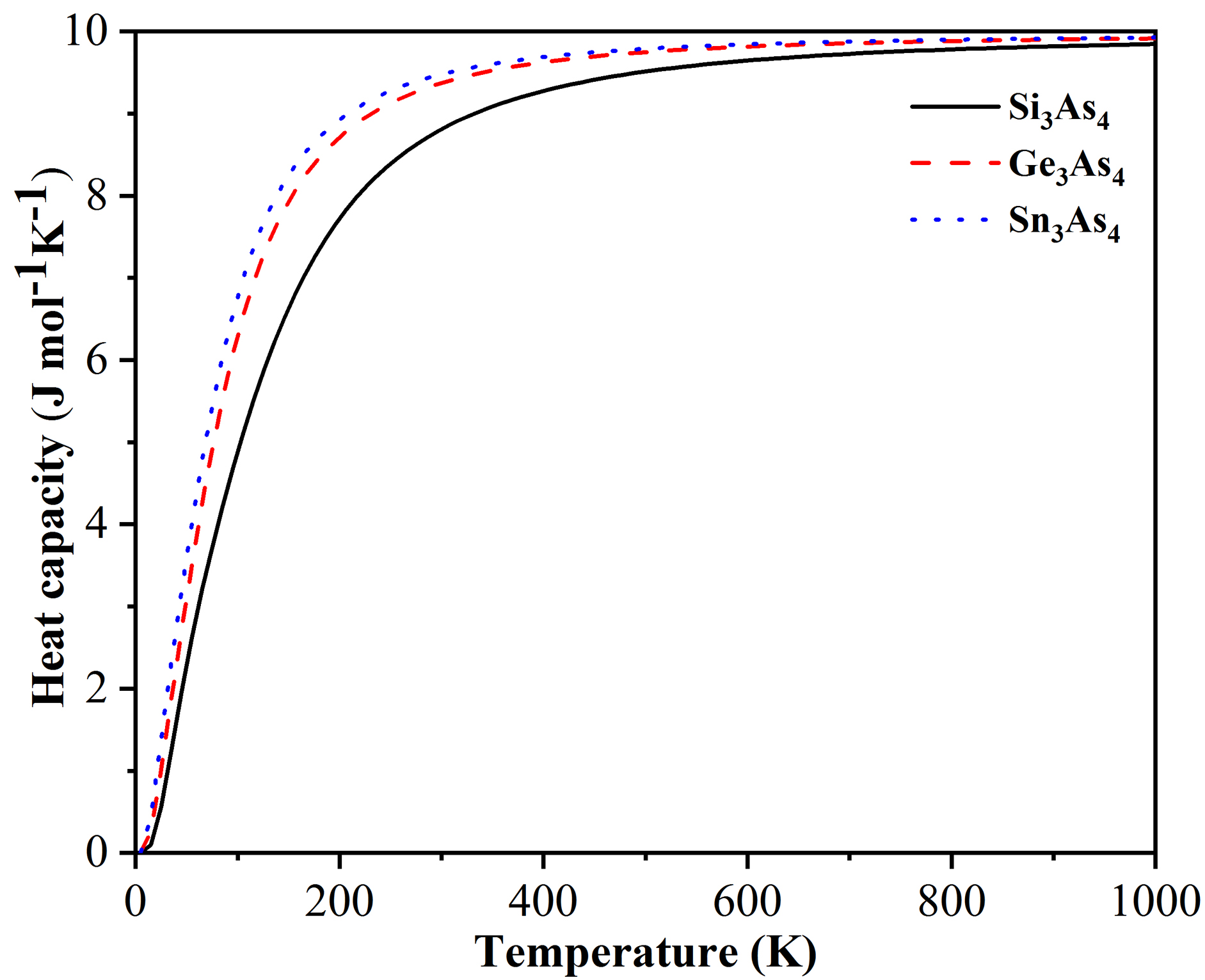}	
\caption{(Colour online) The relationship between the constant volume heat 
capacity and temperature for t-X$_{3}$As$_{4}$.}
\label{fig9}
\end{figure}

\begin{figure}[!t]
	\centering
		\includegraphics[width=2.9in]{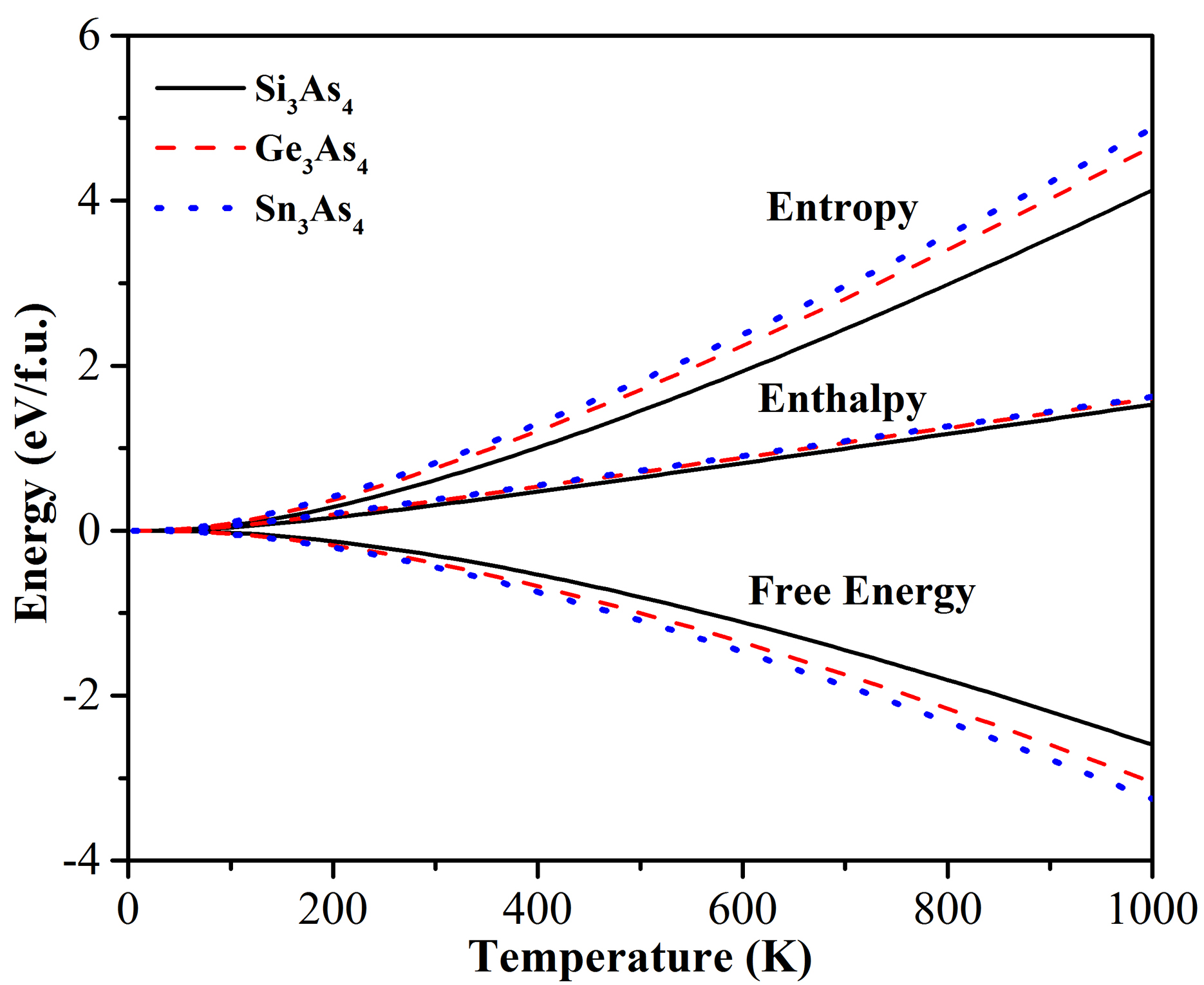}	
\caption{(Colour online) The entropy, the enthalpy and the free energy of 
t-X$_{3}$As$_{4}$ as a function of temperature.}
\label{fig10}
\end{figure}

Heat capacity is an important parameter in condensed matter physics. It can 
also describe the vibrational properties in the heat transition process. The 
lattice (or phonon) contribution and the electron contribution are two main 
sources of heat capacity. The former contributes most at a low temperature, 
while the latter plays an important role at a high temperature. The relations 
of the constant volume heat capacity $C_{V}$ with temperature are presented 
in figure~\ref{fig9} for t-X$_{3}$As$_{4}$. It can be seen that these heat 
capacities increase rapidly with the temperature increase at a lower 
temperature. At a high temperature, heat capacities rise slowly and are close 
to the Dulong-Petit limit. It indicates that the atomic interactions in 
t-X$_{3}$As$_{4}$ occur at a low temperature. The Dulong-Petit limit of 
t-X$_{3}$As$_{4}$ is about 10~J~mol$^{-1}$K$^{-1}$. The variations of the 
entropy, enthalpy and free energy with temperature at 0~Gpa are shown in 
figure~\ref{fig10}. All values are given in the form of per t-X$_{3}$As$_{4}$ 
formula unit. It is noted that the free energy decreases with the temperature 
increase, and the entropy increases more rapidly than that of enthalpy as 
temperature increases. The calculated values of the entropy, enthalpy and 
free energy of t-X$_{3}$As$_{4}$ at room temperature are listed in 
table~\ref{tab6}. All absolute values of the entropy, enthalpy and free 
energy for t-Sn$_{3}$As$_{4}$ are larger than those of t-Si$_{3}$As$_{4}$ 
and t-Ge$_{3}$As$_{4}$ under the temperature range from 0 to 1000~K, which 
means that t-Sn$_{3}$As$_{4}$ are less stable than t-Si$_{3}$As$_{4}$ and 
t-Ge$_{3}$As$_{4}$. From the figure~\ref{fig10} and table~\ref{tab6}, we can 
see that the descending order of thermodynamic stability is from 
t-Si$_{3}$As$_{4}$ to t-Ge$_{3}$As$_{4}$ to t-Sn$_{3}$As$_{4}$.

\begin{table}[!t]
\caption{The entropy, enthalpy and free energy (eV/f.u.) for 
t-X$_{3}$As$_{4}$.}
	\begin{center}
		\begin{tabular}{|c|c|c|c|}
			\hline \hline
			  & Si$_{3}$As$_{4}$& Ge$_{3}$As$_{4}$& Sn$_{3}$As$_{4}$\\
			\hline \hline
			Entropy&0.602&0.744&0.807\\
			\hline
			Enthalpy&0.305&0.355&0.377\\
			\hline
			Free energy&$-$0.298&$-$0.388&$-$0.436\\
			\hline\hline
		\end{tabular}
	\end{center}
	\label{tab6}
\end{table}

\section{Conclusion}

For the novel predicted t-X$_{3}$As$_{4}$, the structural, mechanical, 
electronic, optical and thermodynamic properties are studied by the 
first-principles calculations. It is found that t-X$_{3}$As$_{4}$ are stable 
by elastic constants and phonons analysis. The t-Sn$_{3}$As$_{4}$ shows to me more 
anisotropic than t-Si$_{3}$As$_{4}$ and t-Ge$_{3}$As$_{4. }$ Due to a 
larger mass difference between As and Si atom, there is a larger optical 
band gap in the dispersion curves of t-Si$_{3}$As$_{4}$. By analyzing the 
density of phonon states in t-Si$_{3}$As$_{4}$, t-Ge$_{3}$As$_{4}$ and t-Sn$_{3}$As$_{4}$, we can see that the vibration 
of As atoms is dominant in the frequency range of $0.8{-}6.8$~eV, $0.2{-}6.1$~eV 
and $4.2{-}7.6$~eV, respectively. The vibration of Si and Ge atoms is dominant 
in the band of $9.5{-}13$~eV and $6.2{-}8.8$~eV, respectively. For 
t-Sn$_{3}$As$_{4}$, in the range of $0.2{-}4.0$~eV, the total PHDOS is from Sn 
and As, which shows that Sn and As have the similar vibrational probability. 
The band structures and densities of state show that the t-X$_{3}$As$_{4}$ 
(X $=$ Si and Ge) are indirect band gap semiconductors with narrow band gaps of 
1.438 and 0.952, respectively. The band structure of t-Sn$_{3}$As$_{4}$ 
shows that it is a direct band gap semiconductor. By the analysis of 
electron density difference and Mulliken charge population, it is found that 
X (Si, Ge and Sn) atoms lose electrons, and As atoms acquire electrons. The 
static refractive indices are found to be 3.94, 4.48 and 3.89~eV for 
t-Si$_{3}$As$_{4}$, t-Ge$_{3}$As$_{4}$ and t-Sn$_{3}$As$_{4}$, respectively. The $\varepsilon _1 (0)$ are 15.5, 20.0 
and 15.1~eV for t-Si$_{3}$As$_{4}$, t-Ge$_{3}$As$_{4}$ and t-Sn$_{3}$As$_{4}$, respectively. The Debye temperature of 
t-Si$_{3}$As$_{4}$ is larger than that of t-Ge$_{3}$As$_{4}$ and 
t-Sn$_{3}$As$_{4}$. The Dulong-Petit limit of t-X$_{3}$As$_{4}$ is about 10~J~mol$^{-1}$K$^{-1}$. The thermodynamic stability of t-Si$_{3}$As$_{4}$ is 
higher. 

\section{Acknowledgements}

This work is supported by the Natural Science Basic Research plan in Shanxi 
Province of China [No.~2016JM1026] and supported by the 111 Project 
[B17035]. This work is also supported by Leihua Electronic and 
Technology Research Institute, Aviation Industry Corporation of China (No.~MJZ-2016-S-44).

\newpage

\ukrainianpart

\title{Структурні, механічні, електронні, оптичні і термодинамічні властивості  t-X$_{3}$As$_{4}$ (X $=$ Si, Ge і Sn) з першопринципних розрахунків}
\author{Р. Янг\refaddr{label1}, Ю. Ма\refaddr{label1}, К. Вей \refaddr{label1}, Д. Жанг\refaddr{label2}, І. Жоу\refaddr{label3}}
\addresses{
\addr{label1} Школа фізики та оптоелектронної інженерії, громадський університет у  Сіані, 710071, Китай
\addr{label2} Національний суперкомп'ютерний  центр в  Шеньчжені, Шеньчжень 518055, Китай
\addr{label3} Інститут електроніки та технологічних досліджень Лейхуа, Авіаційна промислова корпорація Китаю,\\ Усі, Цзянсу 214063, Китай
}

\makeukrtitle

\begin{abstract}
Структурні, механічні, електронні, оптичні і термодинамічні властивості  t-X$_{3}$As$_{4}$ (X $=$ Si, Ge і Sn)
з тетрагональною структурою досліджено з першопринципних розрахунків.  Результати обчислень показують, що  ці сполуки є механічно і динамічно стійкими.  Дослідивши пружну анізотропію, встановлено, що анізотропія  t-Sn$_{\mathrm{3}}$As$_{\mathrm{4}}$ є сильніша, ніж анізотропія t-Si$_{\mathrm{3}}$As$_{\mathrm{4}}$ і 
t-Ge$_{\mathrm{3}}$As$_{\mathrm{4}}$. Зонна структура і густина станів показують, що  t-X$_{\mathrm{3}}$As$_{\mathrm{4}}$ (Si, Ge і Sn) --- це напівпровідники  з вузькими забороненими зонами. На основі аналізу різниці електронної густини встановлено, що у t-X$_{\mathrm{3}}$As$_{\mathrm{4}}$ атоми As отримують електрони, а X атоми втрачають електрони. Розраховані статичні діелектричні сталі, $\varepsilon_{1} (0)$, є 15.5, 20.0 і 15.1~еВ відповідно для 
t-X$_{\mathrm{3}}$As$_{\mathrm{4}}$ (X $=$ Si, Ge і Sn). Границя Дюлонга-Пті t-X$_{\mathrm{3}}$As$_{\mathrm{4}}$ є біля  10~Дж$\cdot$моль$^{\mathrm{-1}}$$\cdot$K$^{\mathrm{-1}}$. Термодинамічна стійкість поступово понижується від  t-Si$_{\mathrm{3}}$As$_{\mathrm{4}}$ до 
t-Ge$_{\mathrm{3}}$As$_{\mathrm{4}}$ і до t-Sn$_{\mathrm{3}}$As$_{\mathrm{4}}$.

\keywords t-X$_{3}$As$_{4}$, механічні властивості, оптоелектронні властивості, термодинамічні властивості, першопринципні розрахунки
\end{abstract}

\end{document}